\documentclass[aps,prb,reprint,preprintnumbers,superscriptaddress,amsmath,amssymb,bibnotes,longbibliography]{revtex4-2}

\usepackage{graphicx}
\usepackage{dcolumn}
\usepackage{bm}
\usepackage{color}
\usepackage[colorlinks,linkcolor=blue,anchorcolor=blue,citecolor=blue,urlcolor=blue,filecolor=blue,menucolor=blue,runcolor=blue]{hyperref}
\begin{document}
\title{Nodeless superconductivity in Lu$_{5-x}$Rh$_6$Sn$_{18+x}$ with broken time reversal symmetry}
\author{A. Wang}
\affiliation{Center for Correlated Matter and Department of Physics, Zhejiang University, Hangzhou 310058, China}
\author{Z. Y. Nie}
\affiliation{Center for Correlated Matter and Department of Physics, Zhejiang University, Hangzhou 310058, China}
\author{F. Du}
\affiliation{Center for Correlated Matter and Department of Physics, Zhejiang University, Hangzhou 310058, China}
\author{G. M. Pang}
\affiliation{Center for Correlated Matter and Department of Physics, Zhejiang University, Hangzhou 310058, China}
\author{N. Kase}
\affiliation{Department of Applied Physics, Tokyo University of Science, 6-3-1 Niijuku, Katsushika-ku, Tokyo 125-8585, Japan}
\author{J. Akimitsu}
\affiliation{ Research Institute for Interdisciplinary Science, Okayama University, 3-1-1 Tsushima-naka, Kitaku,  Okayama 700- 8530, Japan}
\author{Y. Chen}
\affiliation{Center for Correlated Matter and Department of Physics, Zhejiang University, Hangzhou 310058, China}
\author{M. J. Gutmann}
\affiliation{ISIS Facility, Rutherford Appleton Laboratory, Chilton, Didcot Oxon OX11 0QX, United Kingdom}
\author{D. T. Adroja}
\affiliation{ISIS Facility, Rutherford Appleton Laboratory, Chilton, Didcot Oxon OX11 0QX, United Kingdom}
\affiliation{Highly Correlated Matter Research Group, Physics Department, University of Johannesburg, P. O. Box 524, Auckland Park 2006, South Africa}
\author{R. S. Perry}
\affiliation{ISIS Facility, Rutherford Appleton Laboratory, Chilton, Didcot Oxon OX11 0QX, United Kingdom}
\affiliation{Centre for Materials Discovery and London Centre for Nanotechnology, University College London, London WC1E 6BT, United Kingdom}
\author{C. Cao}
\affiliation{Condensed Matter Group, Department of Physics, Hangzhou Normal University, Hangzhou 311121, China}
\affiliation{Center for Correlated Matter and Department of Physics, Zhejiang University, Hangzhou 310058, China}
\author{M. Smidman}
\email{msmidman@zju.edu.cn}
\affiliation{Center for Correlated Matter and Department of Physics, Zhejiang University, Hangzhou 310058, China}
\affiliation{Zhejiang Province Key Laboratory of Quantum Technology and Device, Department of Physics, Zhejiang University, Hangzhou  310058, China}
\author{H. Q. Yuan}
\email{hqyuan@zju.edu.cn}
\affiliation{Center for Correlated Matter and Department of Physics, Zhejiang University, Hangzhou 310058, China}
\affiliation{Zhejiang Province Key Laboratory of Quantum Technology and Device, Department of Physics, Zhejiang University, Hangzhou  310058, China}
\affiliation{State Key Laboratory of Silicon Materials, Zhejiang University, Hangzhou 310058, China}
\affiliation{Collaborative Innovation Center of Advanced Microstructures, Nanjing 210093, China}

\date{\today}

\begin{abstract}
Evidence for broken time reversal symmetry (TRS) has been found in the superconducting states of the  $R_5$Rh$_6$Sn$_{18}$ (R = Sc, Y, Lu) compounds with  a centrosymmetric caged  crystal structure, but the origin of this phenomenon is unresolved.  Here we report neutron diffraction measurements of single crystals with $R$=Lu, as well as  measurements of the temperature dependence of the magnetic penetration depth using a self-induced tunnel diode-oscillator (TDO) based technique, together with band structure calculations using density functional theory. Neutron diffraction measurements reveal that the system crystallizes in a tetragonal caged structure, and that one of nominal Lu sites in the Lu$_5$Rh$_6$Sn$_{18}$ structure is occupied by Sn, yielding a composition Lu$_{5-x}$Rh$_6$Sn$_{18+x}$ ($x=1$).  The low temperature penetration depth shift $\Delta\lambda(T)$ exhibits an exponential temperature dependence below around $0.3T_c$, giving clear evidence for fully gapped superconductivity. The derived  superfluid density is reasonably well accounted for by a single gap $s$-wave model, whereas agreement cannot be found for models of TRS breaking states with two-component order parameters. Moreover, band structure calculations reveal multiple bands crossing the Fermi level, and indicate that the aforementioned TRS breaking states would be expected to have nodes on the Fermi surface, in constrast to the observations.

\end{abstract}

\maketitle

\section{INTRODUCTION}

The breaking of time-reversal symmetry (TRS) in the superconducting state is manifested by the spontaneous appearance of magnetic fields below the superconducting transition temperature $T_{\rm c}$. This requires the superconducting order parameter to have multiple components with non-trivial phase differences, and hence TRS breaking is a signature of unconventional superconductivity beyond the $s$-wave pairing state of Bardeen-Cooper-Schrieffer (BCS) theory \cite{Ghosh2020,wysokinski2019}.  TRS breaking in the superconducting state was first reported in a few strongly correlated electron systems, such as U$_{1-x}$Th$_x$Be$_{13}$ \cite{Heffner1990}, UPt$_3$ \cite{Luke1993}, and Sr$_2$RuO$_4$ \cite{Luke1998}. These superconductors have generally been found to have nodal superconducting gaps characteristic of non-$s$-wave superconductivity, which is readily anticipated due to the strong Coulomb repulsion in such strongly correlated systems \cite{Joynt2002,Mackenzie2003}.

On the other hand, in recent years TRS breaking has been reported in several weakly correlated superconductors, a number of which have noncentrosymmetric crystal structures, such as LaNiC$_2$ \cite{Hillier2009}, La$_7$(Ir,Rh)$_3$ \cite{Barker2015,Singh2018a}, several Re-based alloys \cite{Singh2017,Shang2018,ShangPRL2018}, and CaPtAs \cite{Xie2020,Shang2020}. Although noncentrosymmetric superconductors have been predicted to exhibit novel superconducting properties due to the influence of anti-symmetric spin-orbit coupling (ASOC) \cite{BauerNCS,Smidman2017}, the relationship between the breaking of TRS and the lack of inversion symmetry is not determined, and moreover, many of the above systems show other behaviors similar to conventional superconductors, such as fully open superconducting gaps. In the case of LaNiC$_2$, it was shown that the breaking of TRS at $T_{\rm c}$ is incompatible with a significant influence of ASOC \cite{Quintanilla2010}. Moreover, the occurrence of TRS breaking together with two-gap superconductivity in both  LaNiC$_2$ \cite{Chen2013}, as well as centrosymmetric LaNiGa$_2$ \cite{Hillier2012,Weng2016}, was accounted for by an even-parity nonunitary triplet pairing state \cite{Weng2016,Ghosh2019}. Furthermore, the recent findings of the signatures of TRS breaking in centrosymmetric elemental rhenium also suggests that broken inversion symmetry is not essential for this phenomenon in Re-based superconductors \cite{ShangPRL2018}.
 
In this context, it is of particular interest to examine the superconductivity of weakly correlated \textit{centrosymmetric} superconductors exhibiting TRS breaking. One such family of superconductors are the caged compounds $R_5$Rh$_6$Sn$_{18}$ ($R$ = Sc, Y, Lu), which crystallize with the tetragonal space group $I4_1/acd$ \cite{Miraglia1987}, and evidence for TRS breaking is found from muon-spin relaxation ($\mu$SR) measurements \cite{Bhattacharyya2015,Bhattacharyya2015A,Bhattacharyya2018} (although TRS breaking signatures were not observed in another study for $R$=Sc \cite{Feig2020}). Resistivity measurements of all three compounds show semiconducting behavior in the normal state, before becoming superconducting below 5.0 K, 3.0 K and 4.0 K for $R=$~Sc, Y, and Lu, respectively  \cite{Kase2011,Kase2011b}. In the case of  Lu$_5$Rh$_6$Sn$_{18}$, the exponentially activated behavior of the electronic specific heat, negligible residual thermal conductivity, and saturation of the superfluid density derived from transverse-field $\mu$SR, yielded clear evidence for fully gapped superconductivity \cite{Kase2011,Zhang2015a,Bhattacharyya2015}. On the other hand, the proposed possible TRS breaking pairing states for  $R_5$Rh$_6$Sn$_{18}$ from the symmetry analysis corresponded to a singlet $d+id$ state with line nodes or nonunitary triplet pairing with point nodes \cite{Bhattacharyya2015}. As such, it is of particular importance to perform additional detailed studies sensitive to the superconducting gap structure, together with electronic structure calculations, in order to identify the nature of the superconducting order parameter and explain the origin of TRS breaking. 

In this work, we report neutron diffraction and  magnetic penetration depth measurements of  single crystals of Lu$_{5-x}$Rh$_6$Sn$_{18+x}$ (denoted LRS), as well as band structure calculations. The temperature dependence of the magnetic penetration depth shift $\Delta\lambda(T)$ measured using the  tunnel-diode oscillator (TDO) based method exhibits exponentially activated behavior for $T\ll T_c$, indicating a nodeless superconducting gap. The derived normalized superfluid density $\rho_s(T)$ can be reasonably well accounted for by a single gap $s$~wave model, consistent with moderately strong electron-phonon coupling.

\section{EXPERIMENTAL DETAILS}

LRS single crystals were synthesized using a Sn-flux method \cite{Remeika1980}. Room temperature single crystal neutron diffraction measurements were performed using the SXD instrument \cite{Keen2006} at the ISIS pulsed neutron facility at the Rutherford Appleton Laboratory. The composition of the crystals was checked by energy-dispersive x-ray spectroscopy, using a JSM-6610LV SEM with an Oxford Instruments EDS detector. The temperature dependence of the electrical resistivity $\rho(T)$ was measured in a $^4$He system down to 2~K, using the four-probe method. Magnetic susceptibility and magnetization measurements down to 2~K were performed using a superconducting quantum interference device (SQUID) magnetometer (MPMS-5T).

The temperature dependence of the London penetration depth shift $\Delta\lambda(T)=\lambda(T)-\lambda(0)$ was measured using a self-induced tunnel diode-oscillator (TDO)-based method in a $^3$He cryostat, at temperatures down to 0.35~K. The samples were cut into regular cuboids, with dimensions less than $800\times800\times300~\mu$m$^3$. The samples were fixed on a sapphire rod using GE varnish, which was inserted into the coil of the TDO circuit, without contact between the sample and coil. The magnetic field in the coil generated by the current in the circuit is about 2~$\mu$T, which is much less than the  lower critical field ($H_{c1}$) of the sample. The operating frequency of the TDO system is about 7~MHz with a noise level of about 0.1~Hz. $\Delta\lambda(T)$ is proportional to the shift of the resonant frequency of the TDO circuit $\Delta f=f(T)-f(0)$, $\Delta\lambda(T)=G\Delta f(T)$, where  $G$ is determined from the sample and coil geometry \cite{Gfactor}.

Band structure calculations  were performed using density functional theory (DFT), as  implemented in the Vienna Abinitio Simulation Package (VASP) \cite{VASP01,VASP02}. To ensure convergence, we employed plane-wave basis up to $400~eV$ and a $3\times3\times3$-centered K-mesh so that the total energy converges to $1~$meV per cell. The Perdew, Burke, and Ernzerhoff parameterization (PBE) of the generalized gradient approximation (GGA) to the exchange correlation functional and spin-orbit coupling are considered in the calculations \cite{GGA}.

\section{RESULTS}
\begin{figure} [t]
\begin{center}
  \includegraphics[width=\columnwidth]{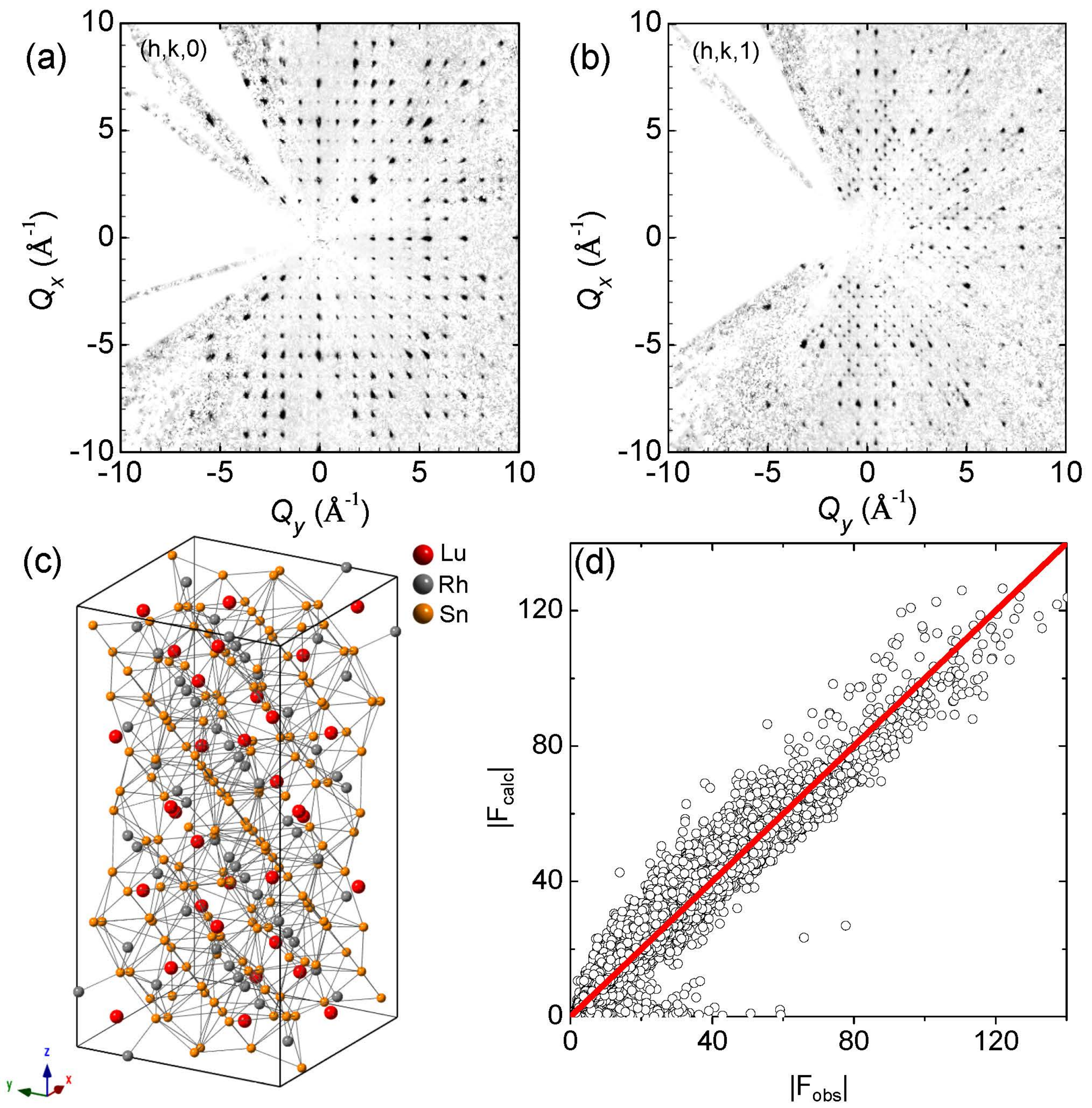}
\end{center}
	\caption{(Color online)  Neutron diffraction measurements are displayed in the (a) $(h k 0)$, and (b) $(h k 1)$ planes, measured on the SXD instrument. (c) Crystal structure of Lu$_{5-x}$Rh$_6$Sn$_{18+x}$ ($x=1$) derived from neutron diffraction measurements, with Lu, Rh, and Sn in red, grey and yellow, respectively. (d) Plot of the calculated ($|F_{\rm calc}|$) vs observed ($|F_{\rm obs}|$) structural factors for the refinement of the neutron diffraction data. }
   \label{Fig1}
\end{figure}

\subsection{Structure}

In order to characterize the crystal structure of LRS, single crystal neutron diffraction  measurements were performed. The results from refining the neutron diffraction data are displayed in Tables~\ref{table:table1} and \ref{table:table2}. Results from the neutron diffraction measurements in the $(h k 0)$ and $(h k 1)$ layers are displayed in Figs.~\ref{Fig1}(a) and (b).  The tetragonal structure with space group $I4_1/acd$  corresponds to the  distortion of a cubic structure (space group $Fm\bar{3}m$), which approximately doubles the length of the $c$-axis, and  is displayed in Fig.~\ref{Fig1}(c). We note that energy-dispersive x-ray spectroscopy measurements of the samples indicate a deficiency of Lu and excess Sn in the crystals, with a composition close to Lu$_4$Rh$_6$Sn$_{19}$. This is in agreement with the structural refinement, where the site with Wyckoff position 8b, which would be occupied by Lu in the nominal Lu$_5$Rh$_6$Sn$_{18}$ structure, is instead entirely occupied by Sn atoms (labelled Sn(7) in Table~\ref{table:table2}). It is clear from the diffraction results that there is a significant  amount of diffuse scattering which is not taken into account of by the above structural model. This can be seen in particular in the diffraction measurements of the $(h k 1)$ layer in Fig.~\ref{Fig1}(b), where the weak superlattice reflections are connected by weak diffuse lines. If the twinning of the crystals is taken into consideration, where the $c$-axis can potentially lie along  one of three orthogonal directions, then this diffuse scattering is well accounted for, with approximately equal occupations for each of the twinned domains \cite{Miraglia1987}, as shown by the plot of the calculated versus observed structure factors for such a refinement in Fig.~\ref{Fig1}(d). This indicates the highly twinned nature of the crystals. 

\begin{table}[tb]
\caption{Parameters and results of the structural refinement for the single crystal neutron  diffraction measurements of  Lu$_{5-x}$Rh$_6$Sn$_{18+x}$.}
\label{table:table1}
\begin{ruledtabular}
\begin{tabular}{l l}
Formula &Lu$_{4}$Rh$_6$Sn$_{19}$ \\
Molar mass (g~mol$^{-1}$)  &3572.8 \\
Wavelength range (\AA) &0.25-8.8 \\
Crystal system &Tetragonal \\
Space group &I4$_1$/acd \\
$a$ (\AA) &13.6696(16) \\
$c$ (\AA) &27.339(3) \\
Cell volume $(\AA^3)$ &5108.50(10) \\
$Z$ &8 \\
Density (calculated) (g~cm$^{-3}$) &9.2908 \\
$F$(000) &1459.16 \\
Crystal size(mm$^3$) &3$\times$5$\times$7 \\
Number of reflections &26533 \\
Data / used$^{\star}$ / $F^2>3\sigma$ / &26533 / 26483 / 21759 / \\
restraints / parameters &0 / 80 \\
Goodness-of-fit on $F^2>3\sigma$/all &3.75/4.05 \\
Final $R_1/wR_2$, $F^2>3\sigma$/all &0.1007/0.2163, 0.1163/0.2208 \\
\end{tabular}
\end{ruledtabular}
{$^{\star}$Outliers with $|F^2_{\rm obs}-F^2_{\rm calc}|>30\sigma(F^2_{\rm obs})$ were omitted from the refinement.}
\end{table}

\begin{table}[tb]
\caption{Atomic positions  for Lu$_{5-x}$Rh$_6$Sn$_{18+x}$ obtained from structural refinements of single crystal neutron diffraction data.}
\label{table:table2}
\begin{ruledtabular}
 \begin{tabular}{c c c c c}
Atom & Wyckoff site &$x$ &$y$ &$z$  \\
\hline\\[-2ex]
Lu(1) &32g &0.63661(12) &0.11648(12) &0.05719(5)\\
Rh(1) &32g &0.67289(19) &0.74143(16) &0.53734(17)\\
Rh(2) &16d &0.5 & 0.75 &0.49828(15)\\
Sn(1) &32g &0.67289(19) &0.74143(16) &0.53734(17)\\
Sn(2) &16f &0.67613(15) &0.92613(15) &0.625\\
Sn(3) &32g &0.6747(2) &0.75567(18) &0.71198(18)\\
Sn(4) &32g &0.58750(14) &0.66168(14) &0.41923(6)\\
Sn(5) &16e &0.21346(12) &0 &0.25\\
Sn(6) &16f &0.67849(15) &0.57151(15) &0.625\\
Sn(7) &8b &0 &0.25 &0.125\\
\end{tabular}
\end{ruledtabular}
\end{table}

\subsection{Physical properties}

\begin{figure} [http!]
\begin{center}
  \includegraphics[width=\columnwidth]{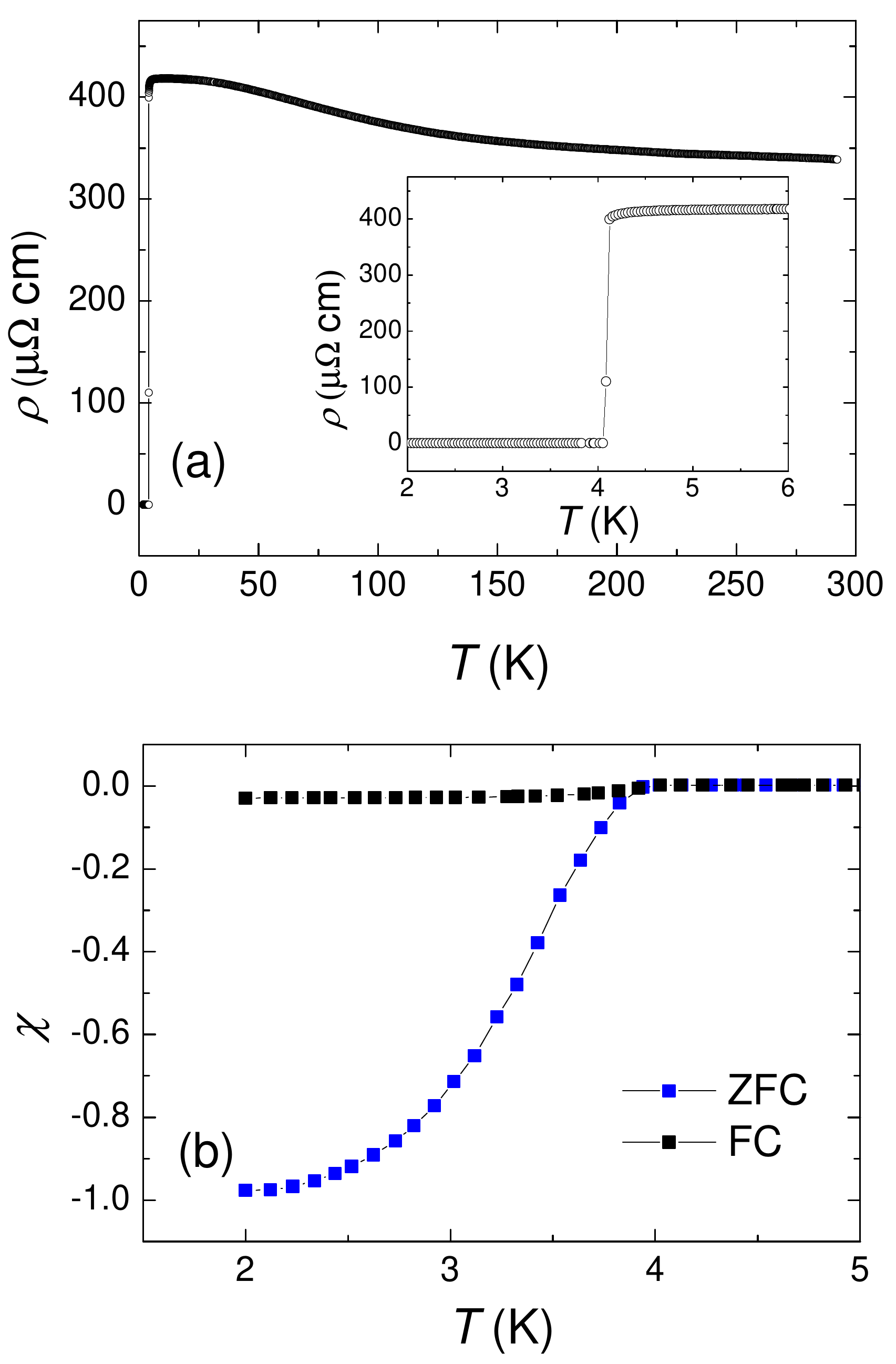}
\end{center}
	\caption{(Color online) (a) Temperature dependence of the electrical resistivity $\rho(T)$ of an LRS single crystal down to 2 K. The inset shows the low temperature part of $\rho(T)$ with a sharp superconducting transition at 4.08 K. (b) Temperature dependence of the magnetic susceptibility, measured upon both zero-field-cooling (ZFC) and field-cooling (FC) in a 10~Oe magnetic field. The data are corrected for demagnetization effects.}
   \label{basic}
\end{figure}

The temperature dependence of the resistivity [$\rho(T)$] of an LRS crystal is displayed in Fig.~\ref{basic}(a), which increases slightly with decreasing temperature, before reaching  a near constant value at low temperatures, in line with previous reports  \cite{Kase2011,Zhang2015a,Bhattacharyya2015}.  A sharp superconducting transition occurs at $T_{\rm c}$= 4.08~K, where $\rho(T)$ drops to zero. A residual resistivity $\rho_0$ of $418~\mu\Omega$~cm is obtained from extrapolating the normal state $\rho(T)$ to zero temperature. The temperature dependence of the magnetic susceptibility is shown in Fig.~\ref{basic}(b) after both zero-field and field-cooling in an applied magnetic field of 10~Oe. A superconducting transition is observed in both quantities, with the onset of diamagnetism occuring at around 3.95~K.

\begin{figure} [http!]
\begin{center}
  \includegraphics[width=\columnwidth]{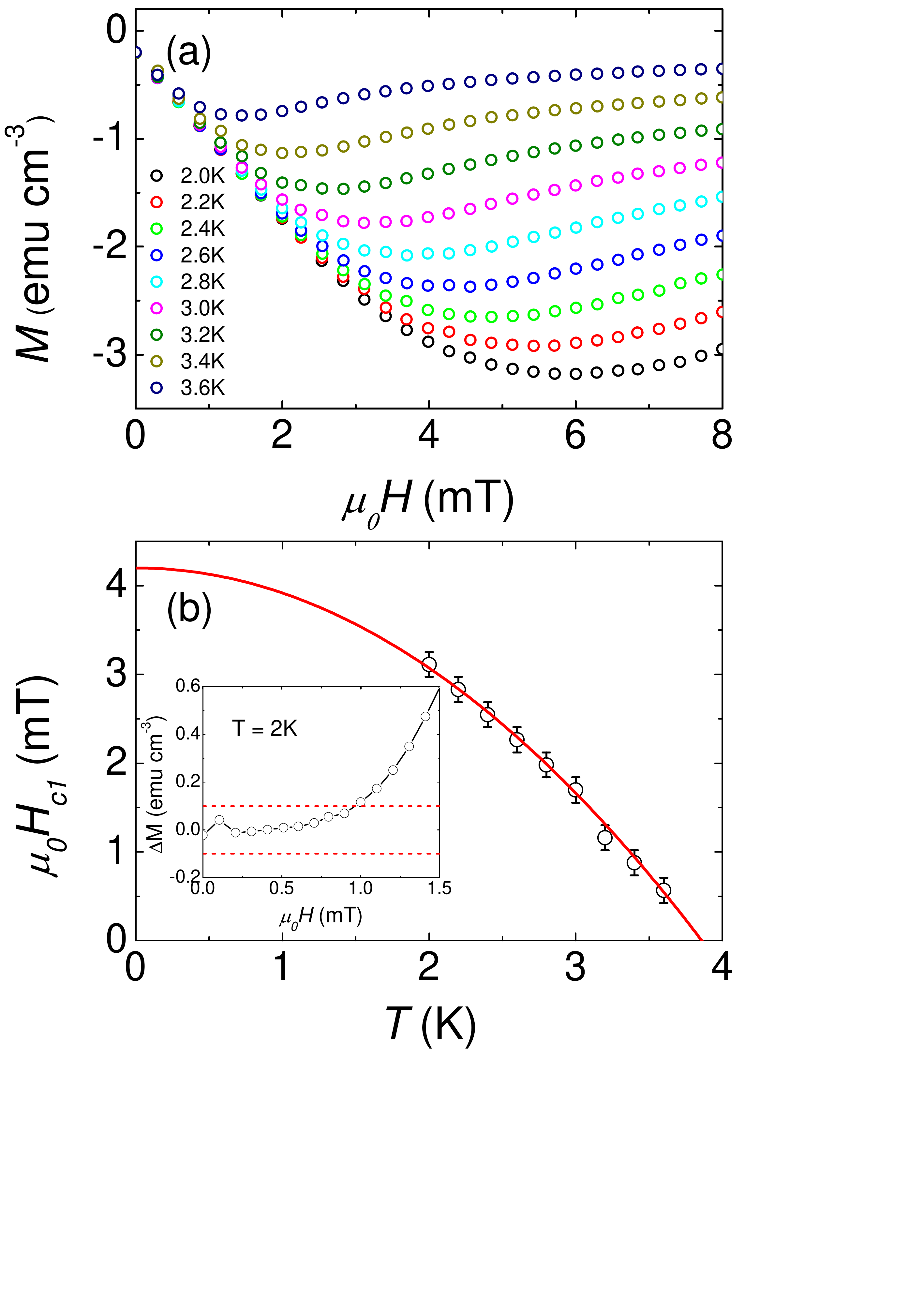}
\end{center}
	\caption{(Color online) (a) Field dependence of the magnetization $M(H)$ of LRS at various temperatures below $T_c$. (b) Temperature dependence of the lower critical field $H_{c1}$ of LRS. The solid red line shows a fit to the Ginzburg-Landau formula. The inset shows the deviation of the magnetization from the low-field linear field dependence at 2~K. The lower critical field was defined as the field where the deviation is greater than $0.1$~emu~cm$^{-3}$.}
   \label{Hc1}
\end{figure}

The field dependence of the magnetization [$M(H)$] is shown in Fig.~\ref{Hc1}(a). The linear decrease of $M(H)$ at low fields indicates that the sample is in the Meissner state. When the applied magnetic field is larger than the lower critical field, there is a deviation from the  low-field linear behavior, due to the sample entering the mixed state. Figure~\ref{Hc1}(b) displays the temperature dependence of the lower critical field $H_{c1}(T)$, which was defined as the  field above which this deviation is larger than   $0.1$~ emu~cm$^{-3}$ [inset of Fig.~\ref{Hc1}(b)], after correcting for demagnetization effects. $H_{c1}(T)$ was fitted using the Ginzburg-Landau formula $H_{c1}(T)=H_{c1}(0)(1-(T/T_c)^2)$ as shown in the main panel of Fig.~\ref{Hc1}(b). A zero temperature value of $\mu_0H_{c1}(0)=4.2$~mT is obtained. Using this value together with an upper critical field of $\mu_0H_{c2}(0)=5.2$~T from Ref.~\onlinecite{Zhang2015a}, we estimated the zero-temperature values of the Ginzburg-Landau coherence length $\xi_{\rm GL}(0)$ and penetration depth $\lambda(0)$ using $\mu_0\mathrm{H}_{c1}=\Phi/(4\pi\lambda^2)\ln{(\lambda/\xi_{\rm GL})}$ and $\mu_0\mathrm{H}_{c2}=\Phi/2\pi\xi_{\rm GL}^2$, which yield $\lambda(0)=390$~nm and  $\xi_{\rm GL}=7.96$~nm.  The mean free path $l$ (in cm) is estimated using $l=1.27\times10^{4}[10^{-6}\rho_0n^{\frac{2}{3}}S/S_{F}]^{-1}$ \cite{Orlando1979}. Using a carrier density $n=1.86\times10^{20}$~cm$^{-3}$ calculated from the value of   $\lambda(0)$, and assuming a spherical Fermi surface with $S/S_{F}\approx1$, the mean free path is estimated to be $9.33~$nm. In the clean and dirty limits, $\xi_{\rm GL}$ is related to the BCS coherence length  $\xi_{\rm BCS}$ via $\xi_{\rm GL}=0.74\xi_{\rm BCS}$ and $\xi_{\rm GL}=0.855(\xi_{\rm BCS}l)^{0.5}$, respectively, yielding respective  $\xi_{\rm BCS}$ of 10.76~nm and 9.28~nm. Both values are comparable to the mean-free path, suggesting that the sample is situated between the clean and dirty limits.

\subsection{Magnetic penetration depth measurements}

\begin{figure} [http!]
\begin{center}
  \includegraphics[width=\columnwidth]{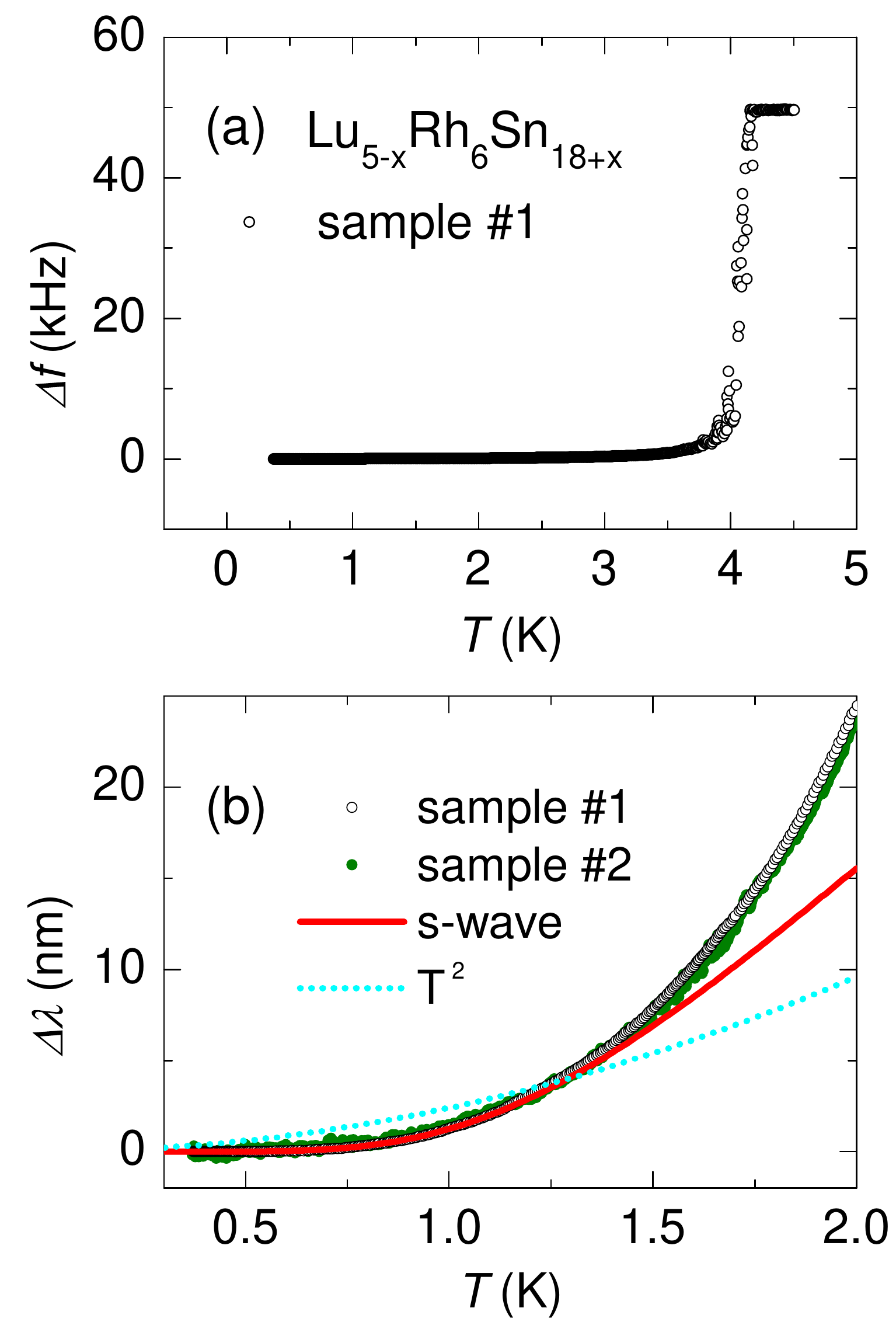}
\end{center}
	\caption{(Color online) (a) Temperature dependence of the resonant frequency shift  $\Delta f(T)$ of a single crystal of  Lu$_{5-x}$Rh$_6$Sn$_{18+x}$ from above $T_{\rm c}$ down to 0.35~K. (b) Low temperature London penetration depth shift $\Delta\lambda(T)$  data for two samples,  where the solid red line shows the fitting to an $s$~wave model, and the dotted blue line shows a quadratic temperature dependence.}
   \label{pd}
\end{figure}

The temperature dependence of the resonant frequency shift  $\Delta f(T)$ from measurements of one sample using the TDO-based method is shown in  Fig.~\ref{pd}(a), while panel (b) displays the low temperature $\Delta\lambda(T)$  of both samples. The calibration factors are $G=4.0~\AA$/Hz and $G=40.0~\AA$/Hz for samples \#1 and \#2 respectively, where the latter value corresponds to a much smaller sample, in order to reduce the effects of sample inhomogeneity. The onset of the superconducting transition in Fig.~\ref{pd}(a) occurs at 4.10 K and ends at about 3.95 K, which is consistent with the resistivity and magnetic susceptibility. It can be seen that $\Delta\lambda(T)$ flattens at the lowest temperatures, being nearly temperature independent,  which indicates a nodeless  superconducting gap. For an isotropic single-band $s$~wave superconductor, $\Delta\lambda(T)$ for $T\ll T_c$ can be approximated using 

\begin{equation}
\Delta\lambda(T)=\lambda(0)\sqrt{\frac{\pi\Delta(0)}{2k_BT}}\textrm{exp}\left(-\frac{\Delta(0)}{k_BT}\right),
\label{equation2}
\end{equation}

\noindent where $\Delta(0)$ is the  superconducting gap magnitude  at zero temperature. This expression gives excellent agreement to the experimental data below $T_c/3$, with  $\Delta(0)=1.5 k_BT_c$. The $\Delta(0)$ is smaller than the value for weak-coupling  BCS theory  of $1.76 k_BT_c$, which can arise from the presence of multiple-gaps and/or gap anisotropy.

\subsection{Superfluid density}

\begin{figure} [http!]
\begin{center}
  \includegraphics[width=\columnwidth]{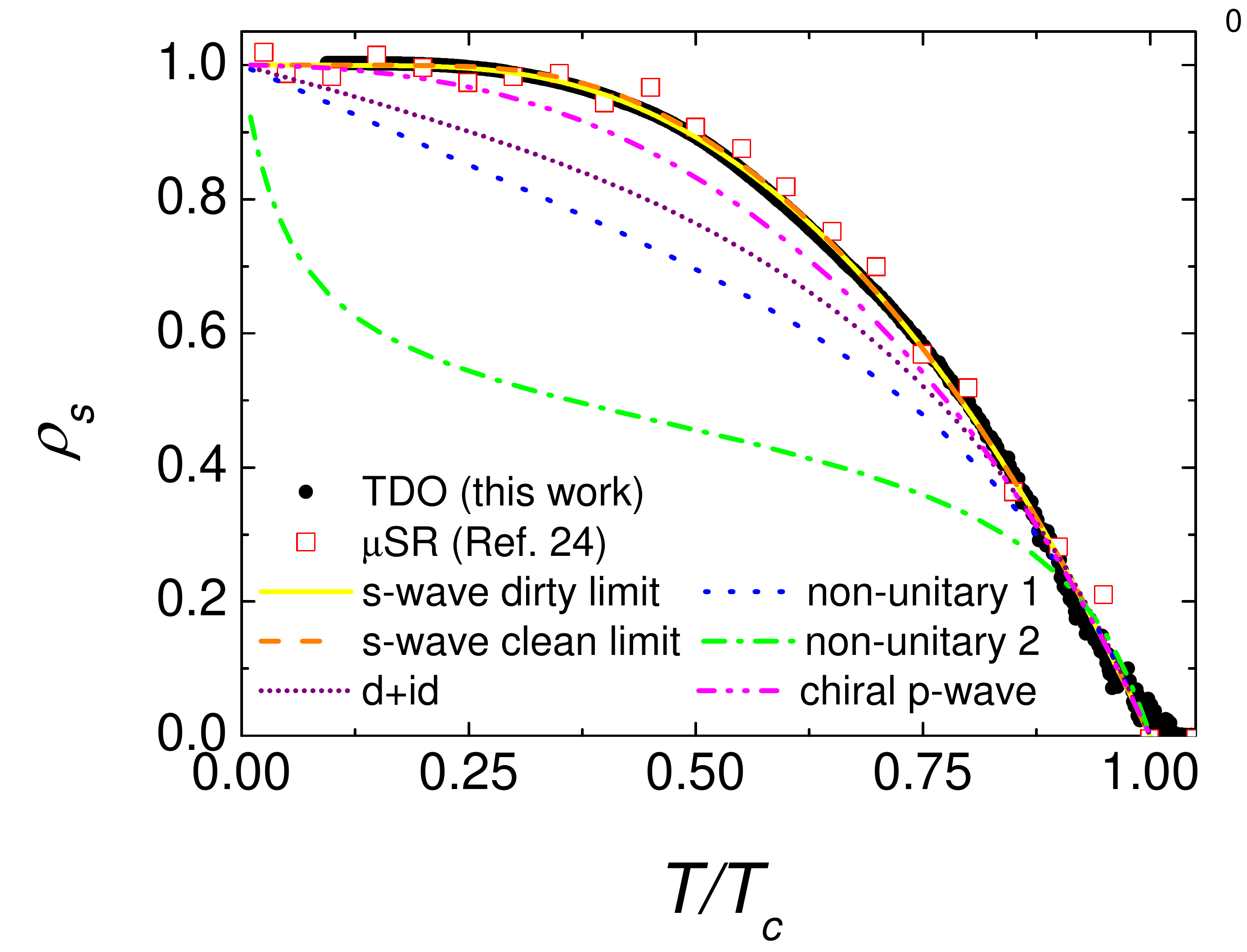}
\end{center}
	\caption{(Color online) Temperature dependence of the normalized superfluid density $\rho_s(T)$ of Lu$_{5-x}$Rh$_6$Sn$_{18+x}$ derived from the TDO-based method. The empty red squares  show the results  from $\mu$SR measurements in Ref.~\onlinecite{Bhattacharyya2015}. The results from fitting with various models are also displayed,  which are described in the text and in Table \ref{table:table3}. }
   \label{rho}
\end{figure}

To determine the gap structure of LRS, the normalized superfluid density was calculated using $\rho_s=[\lambda(0)/\lambda(T)]^2$ for the first sample with $\lambda(0)=390$~nm,  as displayed in Fig.~\ref{rho}. The $\rho_s$ derived from $\mu$SR measurements in Ref.~\onlinecite{Bhattacharyya2015} are also displayed, and it can be seen that there is excellent agreement with the TDO-method results in the present study. The superfluid density were  fitted with various models for the gap structure, and the results are also displayed.  $\rho_s$ was calculated using

\begin{equation}
\rho_s(T)=1+2\left\langle\int_{\Delta_{\overrightarrow{k}}(T)}^{\infty}\frac{E{\rm d}E}{\sqrt{E^2-\Delta_{k}(T)^2}}\frac{\partial f}{\partial E}\right\rangle_{\rm FS},
\label{equation3}
\end{equation}

\noindent where $f(E, T)$ is the Fermi-Dirac function  and $\left\langle\ldots\right\rangle_{\rm FS}$ represents an average over a spherical Fermi surface \cite{Prozorov2006}. The  gap function is given by

\begin{equation}
\Delta_{k}(\mathbf{k},T)=\Delta(T)g_{k}(\mathbf{k}),
\label{equation6}
\end{equation}

\noindent where $g_{k}(\mathbf{k})$ is the angle dependence of the gap function, while the temperature dependence is given by

\begin{equation}
\Delta(T)=\Delta(0){\rm tanh}\left\{1.82\left[1.018\left(T_c/T-1\right)\right]^{0.51}\right\}.
\label{equation4}
\end{equation}

\noindent In the case of anisotropic gap functions, it is important to consider the field direction in the experiment due to the anisotropy of the penetration depth. Since our diffraction results reveal the crystals to be highly twinned, with nearly equal populations of three different domains, we calculated a polycrystalline average of the superfluid density for three different crystallographic directions using $\rho_s=(\sqrt{\rho_{aa}\rho_{bb}}+\sqrt{\rho_{bb}\rho_{cc}}+\sqrt{\rho_{cc}\rho_{aa}})/3$ \cite{Maisuradze2009}, where  $\rho_{aa}$, $\rho_{bb}$ and $\rho_{cc}$ are calculated for a spherical  Fermi surface \cite{Prozorov2006}.

 On the other hand, for an $s$-wave superconductor in the dirty limit, $\rho_s$ is given by \cite{Tinkham2004}

\begin{equation}
\rho_s(T)=\frac{\Delta(T)}{\Delta(0)}{\rm tanh}\left\{\frac{\Delta(T)}{2k_BT_c}\right\},
\label{equation5}
\end{equation}

\begin{table}[tb]
\caption{Different models used to fit the superfluid density of  Lu$_{5-x}$Rh$_6$Sn$_{18+x}$, displayed in Fig.~\ref{rho}, together with the fitted gap values. For the nonunitary triplet state with a two-component order parameter, models with three different values of the real parameters $a$ and $b$ are displayed. For the $d+id$ state, $\Delta(0)$ corresponds to the maximum gap magnitude, while for the nonunitary triplet state it is the gap size in the equatorial plane.  }
\label{table:table3}
\begin{ruledtabular}
 \begin{tabular}{c c c c c c }

 label  &  $g_{k}$  &  $\Delta(0)/k_B T_c$  \\[1ex]
  \hline\\[-1ex]
  s-wave clean & constant & 2.08  \\
  s-wave dirty & constant & 1.82  \\[1ex]
  \hline\\[-1ex]
    & $\frac{\sqrt{a^{2}k_{z}^{2}+b^{2}(k_{x}^{2}+k_{y}^{2})} \pm \sqrt{a^{2}k_{z}^{2}}}{b}$ & \\[1ex]
    nonunitary 1 &  $a=b$  &  1.9  \\[1ex]
    nonunitary 2 &  $a/b=5$  &  0.65  \\[1ex]
    chiral p-wave &  $a=0$  &  2.6  \\[1ex]
  \hline\\[-1ex]
    d+id  & $\sqrt{k_{z}^{2}(k_{x}^{2}+k_{y}^{2})}$ &  3.0  \\[1ex]

\end{tabular}
\end{ruledtabular}
\end{table}

Figure~\ref{rho} displays the results from fitting $\rho_s$ with isotropic $s$-wave models in both the clean and dirty limits, as well as for two TRS breaking states described in Ref.~\onlinecite{Bhattacharyya2015}. The angular dependences of these models, together with the gap values, are listed in Table \ref{table:table3}. It can be seen that the single gap $s$-wave models can well describe the data in both the clean and dirty limits, with respective gap values of $\Delta(0)=2.08k_BT_c$ and $\Delta(0)=1.82k_BT_c$. We note that there is a small discrepancy between these models and the data at low temperatures. Together with a slightly smaller  $\Delta(0)$ being derived from the analysis of the low temperature $\Delta\lambda(T)$ (Fig.~\ref{pd}), this could suggest the presence of  a second superconducting gap, since the low temperature $\Delta\lambda(T)$ would be primarily determined by the smaller of the two gaps. If the data are fitted by a two-gap $s$-wave model, the small gap only has a weighting of 5\%. The origin may also be due to a moderate gap anisotropy, as was inferred for Y$_5$Rh$_6$Sn$_{18}$ \cite{Kase2011}. In this case the low temperature $\Delta\lambda(T)$ will be particularly sensitive to the size of the gap minimum, which may explain the smaller value of $\Delta(0)$.

On the other hand, the models for the TRS breaking states proposed in Ref.~\onlinecite{Bhattacharyya2015} are unable to describe the data. The $d+id$ state has two point nodes at the poles and an equatorial  line node,  while the nonunitary states only have the  two point nodes. Since the nonunitary state has a two-component order parameter, corresponding to the triplet order parameter $\textbf{d(k)} = (AZ, iAZ, B(X+iY))$, there will in general be a two-gap structure $|\Delta_{\pm}(\mathbf{k})|^2=|\mathbf{d}(\mathbf{k})|^2\pm|\mathbf{d}^*(\mathbf{k})\times\mathbf{d}(\mathbf{k})|$, which depends on the relative value of two adjustable parameters $A$ and $B$, and we show the results from fitting this model for three cases. Note that in Table~\ref{table:table3}, $\Delta(0)$ for the nonunitary states corresponds to the gap magnitude in the equatorial plane, which is the same for both gaps. $A=0$ corresponds to the scenario of a unitary chiral $p$-wave state with $\Delta_{+}=\Delta_{-}$, and although this curve is closest to the experimental data out of the three cases, there is still a significant deviation. Upon increasing $A$, the calculated $\rho_s$ drops more rapidly with increasing temperature, which is in contrast to the data. Furthermore,  all  these TRS breaking models have nodes on a spherical Fermi surface, while the data shows an apparent saturation of $\rho_s$ below $T_c/3$, strongly indicating a nodeless superconducting gap in agreement with the analysis of $\Delta\lambda(T)$.

\subsection{Band structure calculations}

\begin{figure*} [t]
\begin{center}
  \includegraphics[width=0.8\textwidth]{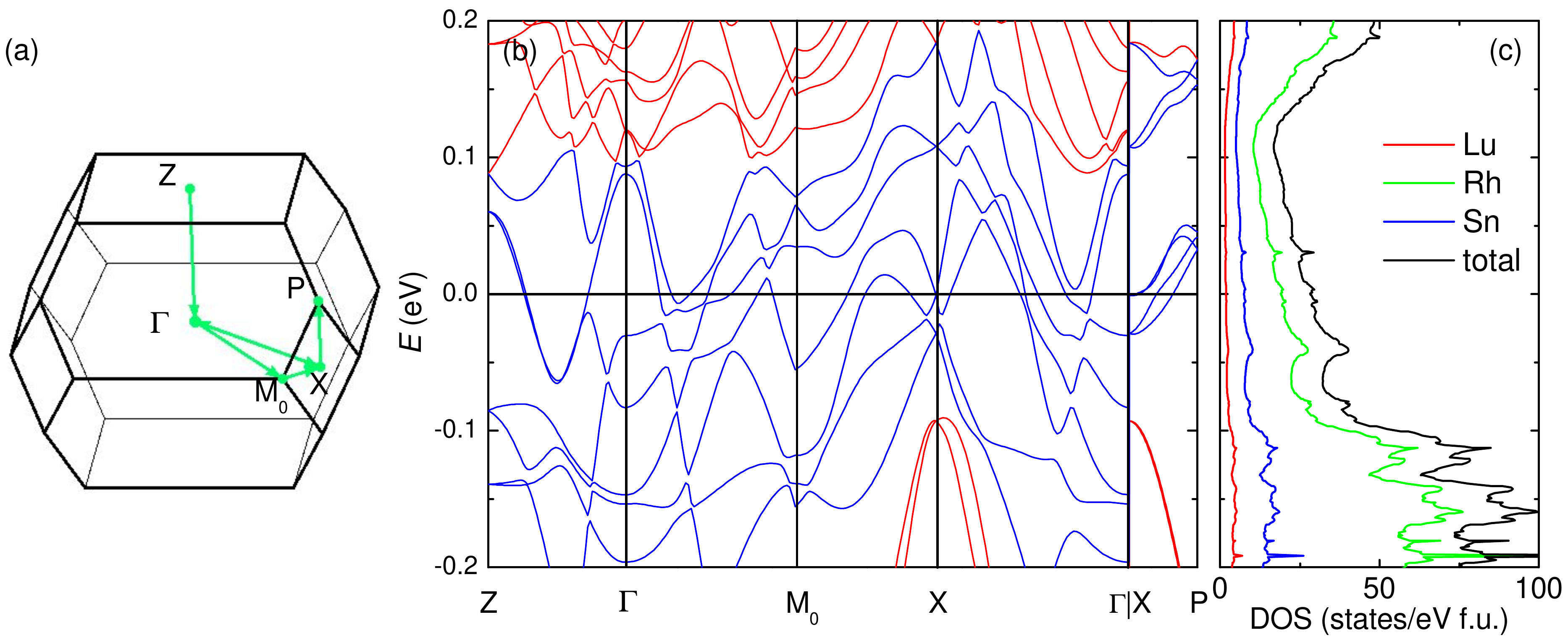}
\end{center}
	\caption{(Color online)  (a) Map of the first Brillouin zone showing the cuts displayed in the plot of the DFT calculations. (b) Band structure of LRS. Blue lines correspond to bands  crossing the Fermi level along the cuts displayed in the panel. The band structure along such a path enables us to confirm the presence of a Fermi surface along both the polar and equatorial directions. (c) Partial and total density of states, obtained from the DFT calculations.}
   \label{band}
\end{figure*}

Figure~\ref{band}(b) displays the band structure of LRS, based on the structural parameters and composition obtained from neutron diffraction and energy-dispersive x-ray spectroscopy, displayed in Tables~\ref{table:table1} and \ref{table:table2}. The band structure is plotted along three high symmetry paths, $\Gamma-Z$, from $X-P$ and a loop in the $ab$ plane passing through $\Gamma$, $M_0$ and $X$ as illustrated in Fig.~\ref{band} (a). Bands that cross the Fermi level are plotted in blue while those which do not are in red. It can be seen that for $\Gamma-Z$ there are two bands which cross the Fermi surface, which indicates that there are two Fermi surface sheets at the poles in the Brillouin zone.  As a result, the time-reversal symmetry breaking states listed in Table~\ref{table:table3} would be expected to have nodes on the Fermi surface, and moreover this indicates that our analysis using a three-dimensional Fermi surface is appropriate. Furthermore, the numerous bands crossing the Fermi level in the $ab$-plane indicate a complex arrangement of Fermi surfaces near the equator. The corresponding total and partial density of states (DOS) are displayed in Fig.~\ref{band}(c). The partial DOS at the Fermi level is $0.277$~states/(eV$\cdot$ atom), $0.559$~states/(eV$\cdot$ atom) and $0.196$~states/(eV$\cdot$ atom) for Lu, Rh and Sn, respectively. Here the largest contributions to the total DOS are from Rh-$4d$ ($61.7\%$), and Sn-$5p$ ($21.4\%$) orbitals, while the contributions from the other orbitals are no larger than $3\%$.

\section{Discussion}

Our findings of nodeless superconductivity in LRS is highly consistent with other studies of the specific heat \cite{Kase2011}, thermal conductivity \cite{Zhang2015a}, and  $\mu$SR \cite{Bhattacharyya2015}. We find that the superfluid density is generally well described by a single gap $s$-wave model, where a small deviation at low temperatures could be due to a moderate gap anisotropy or  multigap superconductivity. In the case of centrosymmetric LaNiGa$_2$, the observation of two-gap superconductivity allowed for the proposal of a nonunitary triplet even parity pairing state to explain the observed TRS breaking \cite{Weng2016}, but whether such a scenario is applicable to LRS remains to be determined. The gap value in the clean limit of $\Delta(0)=2.08k_BT_{\rm c}$ is enhanced over the weak coupling BCS value, which could indicate strong electron-phonon coupling, which is consistent with the specific heat \cite{Kase2011}.

On the other hand, a single-gap $s$-wave model is unable to account for the breaking of TRS inferred from $\mu$SR measurements \cite{Bhattacharyya2015}. We also find that the two previously described TRS breaking pairing states corresponding to two-dimensional irreducible representations of the crystallographic point group are unable to describe the observed $\rho_s$. A possible explanation for this discrepancy could be that the superfluid density in such a model are dominated by the low energy nodal excitations, and therefore if the Fermi surface were absent in the regions of the Brillouin zone where the gap function goes to zero, the superconducting gap would remain fully open over the entire Fermi surface. However, our band structure calculations reveal that there are bands which cross the Fermi level along $\Gamma-Z$, and hence for both TRS breaking pairing states point nodes would be expected to be detected. It is therefore of particular interest to both perform additional measurements to confirm the TRS breaking in LRS such as via Kerr effect measurements \cite{Xia2006}, and to scrutinize other theoretical proposals for reconciling fully gapped superconductivity with broken TRS, such as loop-current order \cite{Ghosh2018}.

\section{Summary}

In summary, we performed single crystal neutron diffraction and tunnel diode oscillator based measurements of the centrosymmetric TRS breaking superconductor   Lu$_{5-x}$Rh$_6$Sn$_{18+x}$. The penetration depth data below $T_c/3$ shows a clear exponential temperature dependence, indicating a nodeless superconducting gap. These findings are corroborated by the analysis of the superfluid density, which are reasonably well described by a single gap $s$-wave model. On the other hand, the data cannot be explained by models with two-component order parameters, which would have nodal gaps on some of the Fermi surface sheets. As a result, further studies are necessary to explain the origin of TRS breaking, and in particular to reconcile this phenomenon with fully gapped superconductivity.

\begin{acknowledgments}
This work was supported by the National Natural Science Foundation of China (No.~11874320,  No.~11974306 and No.~U1632275), the National Key R\&D Program of China (Grants No.~2017YFA0303100 and No.~2016YFA0300202), and the Key R\&D Program of Zhejiang Province, China (2021C01002). We would like to acknowledge access to the equipment to perform EDX analysis at the Research Complex at Harwell.
\end{acknowledgments}


\begin{thebibliography}{41}%
\makeatletter
\providecommand \@ifxundefined [1]{%
 \@ifx{#1\undefined}
}%
\providecommand \@ifnum [1]{%
 \ifnum #1\expandafter \@firstoftwo
 \else \expandafter \@secondoftwo
 \fi
}%
\providecommand \@ifx [1]{%
 \ifx #1\expandafter \@firstoftwo
 \else \expandafter \@secondoftwo
 \fi
}%
\providecommand \natexlab [1]{#1}%
\providecommand \enquote  [1]{``#1''}%
\providecommand \bibnamefont  [1]{#1}%
\providecommand \bibfnamefont [1]{#1}%
\providecommand \citenamefont [1]{#1}%
\providecommand \href@noop [0]{\@secondoftwo}%
\providecommand \href [0]{\begingroup \@sanitize@url \@href}%
\providecommand \@href[1]{\@@startlink{#1}\@@href}%
\providecommand \@@href[1]{\endgroup#1\@@endlink}%
\providecommand \@sanitize@url [0]{\catcode `\\12\catcode `\$12\catcode
  `\&12\catcode `\#12\catcode `\^12\catcode `\_12\catcode `\%12\relax}%
\providecommand \@@startlink[1]{}%
\providecommand \@@endlink[0]{}%
\providecommand \url  [0]{\begingroup\@sanitize@url \@url }%
\providecommand \@url [1]{\endgroup\@href {#1}{\urlprefix }}%
\providecommand \urlprefix  [0]{URL }%
\providecommand \Eprint [0]{\href }%
\providecommand \doibase [0]{https://doi.org/}%
\providecommand \selectlanguage [0]{\@gobble}%
\providecommand \bibinfo  [0]{\@secondoftwo}%
\providecommand \bibfield  [0]{\@secondoftwo}%
\providecommand \translation [1]{[#1]}%
\providecommand \BibitemOpen [0]{}%
\providecommand \bibitemStop [0]{}%
\providecommand \bibitemNoStop [0]{.\EOS\space}%
\providecommand \EOS [0]{\spacefactor3000\relax}%
\providecommand \BibitemShut  [1]{\csname bibitem#1\endcsname}%
\let\auto@bib@innerbib\@empty
\bibitem [{\citenamefont {Ghosh}\ \emph
  {et~al.}(2020{\natexlab{a}})\citenamefont {Ghosh}, \citenamefont {Smidman},
  \citenamefont {Shang}, \citenamefont {Annett}, \citenamefont {Hillier},
  \citenamefont {Quintanilla},\ and\ \citenamefont {Yuan}}]{Ghosh2020}%
  \BibitemOpen
  \bibfield  {author} {\bibinfo {author} {\bibfnamefont {S.}~\bibnamefont
  {Ghosh}}, \bibinfo {author} {\bibfnamefont {M.}~\bibnamefont {Smidman}},
  \bibinfo {author} {\bibfnamefont {T.}~\bibnamefont {Shang}}, \bibinfo
  {author} {\bibfnamefont {J.~F.}\ \bibnamefont {Annett}}, \bibinfo {author}
  {\bibfnamefont {A.~D.}\ \bibnamefont {Hillier}}, \bibinfo {author}
  {\bibfnamefont {J.}~\bibnamefont {Quintanilla}}, and\ \bibinfo {author}
  {\bibfnamefont {H.~Q.}\ \bibnamefont {Yuan}},\ }\bibfield  {title} {\bibinfo
  {title} {Recent progress on superconductors with time-reversal symmetry
  breaking},\ }\href {https://doi.org/10.1088/1361-648X/abaa06}
  {\bibfield  {journal} {\bibinfo  {journal} {J. Phys.: Condens. Matter}\ } \textbf {\bibinfo {volume} {33}},\ \bibinfo
  {pages} {033001} (\bibinfo {year} {2020})}\BibitemShut {NoStop}%
\bibitem [{\citenamefont {Wysoki{\'n}ski}(2019)}]{wysokinski2019}%
  \BibitemOpen
  \bibfield  {author} {\bibinfo {author} {\bibfnamefont {K.~I.}\ \bibnamefont
  {Wysoki{\'n}ski}},\ }\bibfield  {title} {\bibinfo {title} {Time reversal
  symmetry breaking superconductors: {Sr$_2$RuO$_4$} and beyond},\ }\href
  {https://doi.org/10.3390/condmat4020047} {\bibfield  {journal} {\bibinfo
  {journal} {Condensed Matter}\ }\textbf {\bibinfo {volume} {4}},\ \bibinfo
  {pages} {47} (\bibinfo {year} {2019})}\BibitemShut {NoStop}%
\bibitem [{\citenamefont {Heffner}\ \emph {et~al.}(1990)\citenamefont
  {Heffner}, \citenamefont {Smith}, \citenamefont {Willis}, \citenamefont
  {Birrer}, \citenamefont {Baines}, \citenamefont {Gygax}, \citenamefont
  {Hitti}, \citenamefont {Lippelt}, \citenamefont {Ott}, \citenamefont
  {Schenck}, \citenamefont {Knetsch}, \citenamefont {Mydosh},\ and\
  \citenamefont {MacLaughlin}}]{Heffner1990}%
  \BibitemOpen
  \bibfield  {author} {\bibinfo {author} {\bibfnamefont {R.~H.}\ \bibnamefont
  {Heffner}}, \bibinfo {author} {\bibfnamefont {J.~L.}\ \bibnamefont {Smith}},
  \bibinfo {author} {\bibfnamefont {J.~O.}\ \bibnamefont {Willis}}, \bibinfo
  {author} {\bibfnamefont {P.}~\bibnamefont {Birrer}}, \bibinfo {author}
  {\bibfnamefont {C.}~\bibnamefont {Baines}}, \bibinfo {author} {\bibfnamefont
  {F.~N.}\ \bibnamefont {Gygax}}, \bibinfo {author} {\bibfnamefont
  {B.}~\bibnamefont {Hitti}}, \bibinfo {author} {\bibfnamefont
  {E.}~\bibnamefont {Lippelt}}, \bibinfo {author} {\bibfnamefont {H.~R.}\
  \bibnamefont {Ott}}, \bibinfo {author} {\bibfnamefont {A.}~\bibnamefont
  {Schenck}}, \bibinfo {author} {\bibfnamefont {E.~A.}\ \bibnamefont
  {Knetsch}}, \bibinfo {author} {\bibfnamefont {J.~A.}\ \bibnamefont {Mydosh}},
  and\ \bibinfo {author} {\bibfnamefont {D.~E.}\ \bibnamefont {MacLaughlin}},\
  }\bibfield  {title} {\bibinfo {title} {New phase diagram for
  {(U,Th)}{${\mathrm{Be}}_{13}$}: {A} muon-spin-resonance and
  {${\mathrm{H}}_{\mathit{C}1}$} study},\ }\href
  {https://doi.org/10.1103/PhysRevLett.65.2816} {\bibfield  {journal} {\bibinfo
   {journal} {Phys. Rev. Lett.}\ }\textbf {\bibinfo {volume} {65}},\ \bibinfo
  {pages} {2816} (\bibinfo {year} {1990})}\BibitemShut {NoStop}%
\bibitem [{\citenamefont {Luke}\ \emph {et~al.}(1993)\citenamefont {Luke},
  \citenamefont {Keren}, \citenamefont {Le}, \citenamefont {Wu}, \citenamefont
  {Uemura}, \citenamefont {Bonn}, \citenamefont {Taillefer},\ and\
  \citenamefont {Garrett}}]{Luke1993}%
  \BibitemOpen
  \bibfield  {author} {\bibinfo {author} {\bibfnamefont {G.~M.}\ \bibnamefont
  {Luke}}, \bibinfo {author} {\bibfnamefont {A.}~\bibnamefont {Keren}},
  \bibinfo {author} {\bibfnamefont {L.~P.}\ \bibnamefont {Le}}, \bibinfo
  {author} {\bibfnamefont {W.~D.}\ \bibnamefont {Wu}}, \bibinfo {author}
  {\bibfnamefont {Y.~J.}\ \bibnamefont {Uemura}}, \bibinfo {author}
  {\bibfnamefont {D.~A.}\ \bibnamefont {Bonn}}, \bibinfo {author}
  {\bibfnamefont {L.}~\bibnamefont {Taillefer}}, and\ \bibinfo {author}
  {\bibfnamefont {J.~D.}\ \bibnamefont {Garrett}},\ }\bibfield  {title}
  {\bibinfo {title} {Muon spin relaxation in {${\mathrm{UPt}}_{3}$}},\ }\href
  {https://doi.org/10.1103/PhysRevLett.71.1466} {\bibfield  {journal} {\bibinfo
   {journal} {Phys. Rev. Lett.}\ }\textbf {\bibinfo {volume} {71}},\ \bibinfo
  {pages} {1466} (\bibinfo {year} {1993})}\BibitemShut {NoStop}%
\bibitem [{\citenamefont {Luke}\ \emph {et~al.}(1998)\citenamefont {Luke},
  \citenamefont {Fudamoto}, \citenamefont {Kojima}, \citenamefont {Larkin},
  \citenamefont {Merrin}, \citenamefont {Nachumi}, \citenamefont {Uemura},
  \citenamefont {Maeno}, \citenamefont {Mao}, \citenamefont {Mori},
  \citenamefont {Nakamura},\ and\ \citenamefont {Sigrist}}]{Luke1998}%
  \BibitemOpen
  \bibfield  {author} {\bibinfo {author} {\bibfnamefont {G.~M.}\ \bibnamefont
  {Luke}}, \bibinfo {author} {\bibfnamefont {Y.}~\bibnamefont {Fudamoto}},
  \bibinfo {author} {\bibfnamefont {K.~M.}\ \bibnamefont {Kojima}}, \bibinfo
  {author} {\bibfnamefont {M.~I.}\ \bibnamefont {Larkin}}, \bibinfo {author}
  {\bibfnamefont {J.}~\bibnamefont {Merrin}}, \bibinfo {author} {\bibfnamefont
  {B.}~\bibnamefont {Nachumi}}, \bibinfo {author} {\bibfnamefont {Y.~J.}\
  \bibnamefont {Uemura}}, \bibinfo {author} {\bibfnamefont {Y.}~\bibnamefont
  {Maeno}}, \bibinfo {author} {\bibfnamefont {Z.~Q.}\ \bibnamefont {Mao}},
  \bibinfo {author} {\bibfnamefont {Y.}~\bibnamefont {Mori}}, \bibinfo {author}
  {\bibfnamefont {H.}~\bibnamefont {Nakamura}}, and\ \bibinfo {author}
  {\bibfnamefont {M.}~\bibnamefont {Sigrist}},\ }\bibfield  {title} {\bibinfo
  {title} {Time-reversal symmetry-breaking superconductivity in
  {${\mathrm{Sr}}_{2}{\mathrm{RuO}}_{4}$}},\ }\href
  {https://doi.org/10.1038/29038} {\bibfield  {journal} {\bibinfo  {journal}
  {Nature}\ }\textbf {\bibinfo {volume} {394}},\ \bibinfo {pages} {558}
  (\bibinfo {year} {1998})}\BibitemShut {NoStop}%
\bibitem [{\citenamefont {Joynt}\ and\ \citenamefont
  {Taillefer}(2002)}]{Joynt2002}%
  \BibitemOpen
  \bibfield  {author} {\bibinfo {author} {\bibfnamefont {R.}~\bibnamefont
  {Joynt}} and\ \bibinfo {author} {\bibfnamefont {L.}~\bibnamefont
  {Taillefer}},\ }\bibfield  {title} {\bibinfo {title} {The superconducting
  phases of {${\mathrm{UPt}}_{3}$}},\ }\href
  {https://doi.org/10.1103/RevModPhys.74.235} {\bibfield  {journal} {\bibinfo
  {journal} {Rev. Mod. Phys.}\ }\textbf {\bibinfo {volume} {74}},\ \bibinfo
  {pages} {235} (\bibinfo {year} {2002})}\BibitemShut {NoStop}%
\bibitem [{\citenamefont {Mackenzie}\ and\ \citenamefont
  {Maeno}(2003)}]{Mackenzie2003}%
  \BibitemOpen
  \bibfield  {author} {\bibinfo {author} {\bibfnamefont {A.}~\bibnamefont
  {Mackenzie}} and\ \bibinfo {author} {\bibfnamefont {Y.}~\bibnamefont
  {Maeno}},\ }\bibfield  {title} {\bibinfo {title} {The superconductivity of
  {${\mathrm{Sr}}_{2}{\mathrm{RuO}}_{4}$} and the physics of spin-triplet
  pairing},\ }\href {https://doi.org/10.1103/RevModPhys.75.657} {\bibfield
  {journal} {\bibinfo  {journal} {Rev. Mod. Phys.}\ }\textbf {\bibinfo {volume}
  {75}},\ \bibinfo {pages} {657} (\bibinfo {year} {2003})}\BibitemShut
  {NoStop}%
\bibitem [{\citenamefont {Hillier}\ \emph {et~al.}(2009)\citenamefont
  {Hillier}, \citenamefont {Quintanilla},\ and\ \citenamefont
  {Cywinski}}]{Hillier2009}%
  \BibitemOpen
  \bibfield  {author} {\bibinfo {author} {\bibfnamefont {A.~D.}\ \bibnamefont
  {Hillier}}, \bibinfo {author} {\bibfnamefont {J.}~\bibnamefont
  {Quintanilla}}, and\ \bibinfo {author} {\bibfnamefont {R.}~\bibnamefont
  {Cywinski}},\ }\bibfield  {title} {\bibinfo {title} {Evidence for
  time-reversal symmetry breaking in the noncentrosymmetric superconductor
  {${\mathrm{LaNiC}}_{2}$}},\ }\href
  {https://doi.org/10.1103/PhysRevLett.102.117007} {\bibfield  {journal}
  {\bibinfo  {journal} {Phys. Rev. Lett.}\ }\textbf {\bibinfo {volume} {102}},\
  \bibinfo {pages} {117007} (\bibinfo {year} {2009})}\BibitemShut {NoStop}%
\bibitem [{\citenamefont {Barker}\ \emph {et~al.}(2015)\citenamefont {Barker},
  \citenamefont {Singh}, \citenamefont {Thamizhavel}, \citenamefont {Hillier},
  \citenamefont {Lees}, \citenamefont {Balakrishnan}, \citenamefont {Paul},\
  and\ \citenamefont {Singh}}]{Barker2015}%
  \BibitemOpen
  \bibfield  {author} {\bibinfo {author} {\bibfnamefont {J.~A.~T.}\
  \bibnamefont {Barker}}, \bibinfo {author} {\bibfnamefont {D.}~\bibnamefont
  {Singh}}, \bibinfo {author} {\bibfnamefont {A.}~\bibnamefont {Thamizhavel}},
  \bibinfo {author} {\bibfnamefont {A.~D.}\ \bibnamefont {Hillier}}, \bibinfo
  {author} {\bibfnamefont {M.~R.}\ \bibnamefont {Lees}}, \bibinfo {author}
  {\bibfnamefont {G.}~\bibnamefont {Balakrishnan}}, \bibinfo {author}
  {\bibfnamefont {D.~M.}\ \bibnamefont {Paul}}, and\ \bibinfo {author}
  {\bibfnamefont {R.~P.}\ \bibnamefont {Singh}},\ }\bibfield  {title} {\bibinfo
  {title} {Unconventional superconductivity in
  {${\mathrm{La}}_{7}{\mathrm{Ir}}_{3}$} revealed by muon spin relaxation:
  Introducing a new family of noncentrosymmetric superconductor that breaks
  time-reversal symmetry},\ }\href
  {https://doi.org/10.1103/PhysRevLett.115.267001} {\bibfield  {journal}
  {\bibinfo  {journal} {Phys. Rev. Lett.}\ }\textbf {\bibinfo {volume} {115}},\
  \bibinfo {pages} {267001} (\bibinfo {year} {2015})}\BibitemShut {NoStop}%
\bibitem [{\citenamefont {{Singh}}\ \emph {et~al.}(2018)\citenamefont
  {{Singh}}, \citenamefont {{Scheurer}}, \citenamefont {{Hillier}},\ and\
  \citenamefont {{Singh}}}]{Singh2018a}%
  \BibitemOpen
  \bibfield  {author} {\bibinfo {author} {\bibfnamefont {D.}~\bibnamefont
  {{Singh}}}, \bibinfo {author} {\bibfnamefont {M.~S.}\ \bibnamefont
  {{Scheurer}}}, \bibinfo {author} {\bibfnamefont {A.~D.}\ \bibnamefont
  {{Hillier}}}, and\ \bibinfo {author} {\bibfnamefont {R.~P.}\ \bibnamefont
  {{Singh}}},\ }\bibfield  {title} {\bibinfo {title} {{Time-reversal-symmetry
  breaking and unconventional pairing in the noncentrosymmetric superconductor
  La$_7$Rh$_3$ probed by $\mu$SR}},,\ }\href
  {https://doi.org/10.1103/PhysRevB.102.134511} {\bibfield  {journal} {\bibinfo
  {journal} {Phys. Rev. B}\ }\textbf {\bibinfo {volume} {102}},\ \bibinfo
  {pages} {134511} (\bibinfo {year} {2020}{\natexlab{a}})}\BibitemShut
  {NoStop}%
\bibitem [{\citenamefont {Singh}\ \emph {et~al.}(2017)\citenamefont {Singh},
  \citenamefont {Barker}, \citenamefont {Thamizhavel}, \citenamefont {Paul},
  \citenamefont {Hillier},\ and\ \citenamefont {Singh}}]{Singh2017}%
  \BibitemOpen
  \bibfield  {author} {\bibinfo {author} {\bibfnamefont {D.}~\bibnamefont
  {Singh}}, \bibinfo {author} {\bibfnamefont {J.~A.~T.}\ \bibnamefont
  {Barker}}, \bibinfo {author} {\bibfnamefont {A.}~\bibnamefont {Thamizhavel}},
  \bibinfo {author} {\bibfnamefont {D.~M.}\ \bibnamefont {Paul}}, \bibinfo
  {author} {\bibfnamefont {A.~D.}\ \bibnamefont {Hillier}}, and\ \bibinfo
  {author} {\bibfnamefont {R.~P.}\ \bibnamefont {Singh}},\ }\bibfield  {title}
  {\bibinfo {title} {Time-reversal symmetry breaking in the noncentrosymmetric
  superconductor {${\mathrm{Re}}_{6}\mathrm{Hf}$: F}urther evidence for
  unconventional behavior in the {$\ensuremath{\alpha}$-Mn} family of
  materials},\ }\href {https://doi.org/10.1103/PhysRevB.96.180501} {\bibfield
  {journal} {\bibinfo  {journal} {Phys. Rev. B}\ }\textbf {\bibinfo {volume}
  {96}},\ \bibinfo {pages} {180501(R)} (\bibinfo {year} {2017})}\BibitemShut
  {NoStop}%
\bibitem [{\citenamefont {Shang}\ \emph
  {et~al.}(2018{\natexlab{a}})\citenamefont {Shang}, \citenamefont {Pang},
  \citenamefont {Baines}, \citenamefont {Jiang}, \citenamefont {Xie},
  \citenamefont {Wang}, \citenamefont {Medarde}, \citenamefont {Pomjakushina},
  \citenamefont {Shi}, \citenamefont {Mesot}, \citenamefont {Yuan},\ and\
  \citenamefont {Shiroka}}]{Shang2018}%
  \BibitemOpen
  \bibfield  {author} {\bibinfo {author} {\bibfnamefont {T.}~\bibnamefont
  {Shang}}, \bibinfo {author} {\bibfnamefont {G.~M.}\ \bibnamefont {Pang}},
  \bibinfo {author} {\bibfnamefont {C.}~\bibnamefont {Baines}}, \bibinfo
  {author} {\bibfnamefont {W.~B.}\ \bibnamefont {Jiang}}, \bibinfo {author}
  {\bibfnamefont {W.}~\bibnamefont {Xie}}, \bibinfo {author} {\bibfnamefont
  {A.}~\bibnamefont {Wang}}, \bibinfo {author} {\bibfnamefont {M.}~\bibnamefont
  {Medarde}}, \bibinfo {author} {\bibfnamefont {E.}~\bibnamefont
  {Pomjakushina}}, \bibinfo {author} {\bibfnamefont {M.}~\bibnamefont {Shi}},
  \bibinfo {author} {\bibfnamefont {J.}~\bibnamefont {Mesot}}, \bibinfo
  {author} {\bibfnamefont {H.~Q.}\ \bibnamefont {Yuan}}, and\ \bibinfo {author}
  {\bibfnamefont {T.}~\bibnamefont {Shiroka}},\ }\bibfield  {title} {\bibinfo
  {title} {Nodeless superconductivity and time-reversal symmetry breaking in
  the noncentrosymmetric superconductor ${{\rm Re}_{24}{\rm Ti}_{5}}$},\ }\href
  {https://doi.org/10.1103/PhysRevB.97.020502} {\bibfield  {journal} {\bibinfo
  {journal} {Phys. Rev. B}\ }\textbf {\bibinfo {volume} {97}},\ \bibinfo
  {pages} {020502(R)} (\bibinfo {year} {2018}{\natexlab{a}})}\BibitemShut
  {NoStop}%
\bibitem [{\citenamefont {Shang}\ \emph
  {et~al.}(2018{\natexlab{b}})\citenamefont {Shang}, \citenamefont {Smidman},
  \citenamefont {Ghosh}, \citenamefont {Baines}, \citenamefont {Chang},
  \citenamefont {Gawryluk}, \citenamefont {Barker}, \citenamefont {Singh},
  \citenamefont {Paul}, \citenamefont {Balakrishnan}, \citenamefont
  {Pomjakushina}, \citenamefont {Shi}, \citenamefont {Medarde}, \citenamefont
  {Hillier}, \citenamefont {Yuan}, \citenamefont {Quintanilla}, \citenamefont
  {Mesot},\ and\ \citenamefont {Shiroka}}]{ShangPRL2018}%
  \BibitemOpen
  \bibfield  {author} {\bibinfo {author} {\bibfnamefont {T.}~\bibnamefont
  {Shang}}, \bibinfo {author} {\bibfnamefont {M.}~\bibnamefont {Smidman}},
  \bibinfo {author} {\bibfnamefont {S.~K.}\ \bibnamefont {Ghosh}}, \bibinfo
  {author} {\bibfnamefont {C.}~\bibnamefont {Baines}}, \bibinfo {author}
  {\bibfnamefont {L.~J.}\ \bibnamefont {Chang}}, \bibinfo {author}
  {\bibfnamefont {D.~J.}\ \bibnamefont {Gawryluk}}, \bibinfo {author}
  {\bibfnamefont {J.~A.~T.}\ \bibnamefont {Barker}}, \bibinfo {author}
  {\bibfnamefont {R.~P.}\ \bibnamefont {Singh}}, \bibinfo {author}
  {\bibfnamefont {D.~M.}\ \bibnamefont {Paul}}, \bibinfo {author}
  {\bibfnamefont {G.}~\bibnamefont {Balakrishnan}}, \bibinfo {author}
  {\bibfnamefont {E.}~\bibnamefont {Pomjakushina}}, \bibinfo {author}
  {\bibfnamefont {M.}~\bibnamefont {Shi}}, \bibinfo {author} {\bibfnamefont
  {M.}~\bibnamefont {Medarde}}, \bibinfo {author} {\bibfnamefont {A.~D.}\
  \bibnamefont {Hillier}}, \bibinfo {author} {\bibfnamefont {H.~Q.}\
  \bibnamefont {Yuan}}, \bibinfo {author} {\bibfnamefont {J.}~\bibnamefont
  {Quintanilla}}, \bibinfo {author} {\bibfnamefont {J.}~\bibnamefont {Mesot}},
  and\ \bibinfo {author} {\bibfnamefont {T.}~\bibnamefont {Shiroka}},\
  }\bibfield  {title} {\bibinfo {title} {Time-reversal symmetry breaking in
  {Re}-based superconductors},\ }\href
  {https://doi.org/10.1103/PhysRevLett.121.257002} {\bibfield  {journal}
  {\bibinfo  {journal} {Phys. Rev. Lett.}\ }\textbf {\bibinfo {volume} {121}},\
  \bibinfo {pages} {257002} (\bibinfo {year} {2018}{\natexlab{b}})}\BibitemShut
  {NoStop}%
\bibitem [{\citenamefont {Xie}\ \emph {et~al.}(2020)\citenamefont {Xie},
  \citenamefont {Zhang}, \citenamefont {Shen}, \citenamefont {Jiang},
  \citenamefont {Pang}, \citenamefont {Shang}, \citenamefont {Cao},
  \citenamefont {Smidman},\ and\ \citenamefont {Yuan}}]{Xie2020}%
  \BibitemOpen
  \bibfield  {author} {\bibinfo {author} {\bibfnamefont {W.}~\bibnamefont
  {Xie}}, \bibinfo {author} {\bibfnamefont {P.}~\bibnamefont {Zhang}}, \bibinfo
  {author} {\bibfnamefont {B.}~\bibnamefont {Shen}}, \bibinfo {author}
  {\bibfnamefont {W.}~\bibnamefont {Jiang}}, \bibinfo {author} {\bibfnamefont
  {G.}~\bibnamefont {Pang}}, \bibinfo {author} {\bibfnamefont {T.}~\bibnamefont
  {Shang}}, \bibinfo {author} {\bibfnamefont {C.}~\bibnamefont {Cao}}, \bibinfo
  {author} {\bibfnamefont {M.}~\bibnamefont {Smidman}}, and\ \bibinfo {author}
  {\bibfnamefont {H.~Q.}~\bibnamefont {Yuan}},\ }\bibfield  {title} {\bibinfo
  {title} {{CaPtAs: A} new noncentrosymmetric superconductor},\ }\href
  {https://doi.org/10.1007/s11433-019-1488-5} {\bibfield  {journal} {\bibinfo
  {journal} {Science China Physics, Mechanics {\&} Astronomy}\ }\textbf
  {\bibinfo {volume} {63}},\ \bibinfo {pages} {237412} (\bibinfo {year}
  {2020})}\BibitemShut {NoStop}%
\bibitem [{\citenamefont {Shang}\ \emph {et~al.}(2020)\citenamefont {Shang},
  \citenamefont {Smidman}, \citenamefont {Wang}, \citenamefont {Chang},
  \citenamefont {Baines}, \citenamefont {Lee}, \citenamefont {Nie},
  \citenamefont {Pang}, \citenamefont {Xie}, \citenamefont {Jiang},
  \citenamefont {Shi}, \citenamefont {Medarde}, \citenamefont {Shiroka},\ and\
  \citenamefont {Yuan}}]{Shang2020}%
  \BibitemOpen
  \bibfield  {author} {\bibinfo {author} {\bibfnamefont {T.}~\bibnamefont
  {Shang}}, \bibinfo {author} {\bibfnamefont {M.}~\bibnamefont {Smidman}},
  \bibinfo {author} {\bibfnamefont {A.}~\bibnamefont {Wang}}, \bibinfo {author}
  {\bibfnamefont {L.-J.}\ \bibnamefont {Chang}}, \bibinfo {author}
  {\bibfnamefont {C.}~\bibnamefont {Baines}}, \bibinfo {author} {\bibfnamefont
  {M.~K.}\ \bibnamefont {Lee}}, \bibinfo {author} {\bibfnamefont {Z.~Y.}\
  \bibnamefont {Nie}}, \bibinfo {author} {\bibfnamefont {G.~M.}\ \bibnamefont
  {Pang}}, \bibinfo {author} {\bibfnamefont {W.}~\bibnamefont {Xie}}, \bibinfo
  {author} {\bibfnamefont {W.~B.}\ \bibnamefont {Jiang}}, \bibinfo {author}
  {\bibfnamefont {M.}~\bibnamefont {Shi}}, \bibinfo {author} {\bibfnamefont
  {M.}~\bibnamefont {Medarde}}, \bibinfo {author} {\bibfnamefont
  {T.}~\bibnamefont {Shiroka}}, and\ \bibinfo {author} {\bibfnamefont {H.~Q.}\
  \bibnamefont {Yuan}},\ }\bibfield  {title} {\bibinfo {title} {Simultaneous
  nodal superconductivity and time-reversal symmetry breaking in the
  noncentrosymmetric superconductor {CaPtAs}},\ }\href
  {https://doi.org/10.1103/PhysRevLett.124.207001} {\bibfield  {journal}
  {\bibinfo  {journal} {Phys. Rev. Lett.}\ }\textbf {\bibinfo {volume} {124}},\
  \bibinfo {pages} {207001} (\bibinfo {year} {2020})}\BibitemShut {NoStop}%
\bibitem [{\citenamefont {Bauer}\ and\ \citenamefont
  {Sigrist}(2012)}]{BauerNCS}%
  \BibitemOpen
  \bibfield  {author} {\bibinfo {author} {\bibfnamefont {E.}~\bibnamefont
  {Bauer}} and\ \bibinfo {author} {\bibfnamefont {M.}~\bibnamefont {Sigrist}},\
  }\href {http://books.google.co.uk/books?id=nDZ4lKD00t8C} {\emph {\bibinfo
  {title} {Non-Centrosymmetric Superconductors: Introduction and Overview}}},\
  Lecture notes in physics\ (\bibinfo  {publisher} {Springer-Verlag Berlin
  Heidelberg},\ \bibinfo {year} {2012})\BibitemShut {NoStop}%
\bibitem [{\citenamefont {Smidman}\ \emph {et~al.}(2017)\citenamefont
  {Smidman}, \citenamefont {Salamon}, \citenamefont {Yuan},\ and\ \citenamefont
  {Agterberg}}]{Smidman2017}%
  \BibitemOpen
  \bibfield  {author} {\bibinfo {author} {\bibfnamefont {M.}~\bibnamefont
  {Smidman}}, \bibinfo {author} {\bibfnamefont {M.~B.}\ \bibnamefont
  {Salamon}}, \bibinfo {author} {\bibfnamefont {H.~Q.}\ \bibnamefont {Yuan}},
  and\ \bibinfo {author} {\bibfnamefont {D.~F.}\ \bibnamefont {Agterberg}},\
  }\bibfield  {title} {\bibinfo {title} {Superconductivity and spin-orbit
  coupling in non-centrosymmetric materials: a review},\ }\href
  {https://doi.org/10.1088/1361-6633/80/3/036501} {\bibfield  {journal}
  {\bibinfo  {journal} {Rep. Prog. Phys.}\ }\textbf {\bibinfo {volume} {80}},\
  \bibinfo {pages} {036501} (\bibinfo {year} {2017})}\BibitemShut {NoStop}%
\bibitem [{\citenamefont {Quintanilla}\ \emph {et~al.}(2010)\citenamefont
  {Quintanilla}, \citenamefont {Hillier}, \citenamefont {Annett},\ and\
  \citenamefont {Cywinski}}]{Quintanilla2010}%
  \BibitemOpen
  \bibfield  {author} {\bibinfo {author} {\bibfnamefont {J.}~\bibnamefont
  {Quintanilla}}, \bibinfo {author} {\bibfnamefont {A.~D.}\ \bibnamefont
  {Hillier}}, \bibinfo {author} {\bibfnamefont {J.~F.}\ \bibnamefont {Annett}},
  and\ \bibinfo {author} {\bibfnamefont {R.}~\bibnamefont {Cywinski}},\
  }\bibfield  {title} {\bibinfo {title} {Relativistic analysis of the pairing
  symmetry of the noncentrosymmetric superconductor {${\text{LaNiC}}_{2}$}},\
  }\href {https://doi.org/10.1103/PhysRevB.82.174511} {\bibfield  {journal}
  {\bibinfo  {journal} {Phys. Rev. B}\ }\textbf {\bibinfo {volume} {82}},\
  \bibinfo {pages} {174511} (\bibinfo {year} {2010})}\BibitemShut {NoStop}%
\bibitem [{\citenamefont {Chen}\ \emph {et~al.}(2013)\citenamefont {Chen},
  \citenamefont {Jiao}, \citenamefont {Zhang}, \citenamefont {Chen},
  \citenamefont {Yang}, \citenamefont {Nicklas}, \citenamefont {Steglich},\
  and\ \citenamefont {Yuan}}]{Chen2013}%
  \BibitemOpen
  \bibfield  {author} {\bibinfo {author} {\bibfnamefont {J.}~\bibnamefont
  {Chen}}, \bibinfo {author} {\bibfnamefont {L.}~\bibnamefont {Jiao}}, \bibinfo
  {author} {\bibfnamefont {J.~L.}\ \bibnamefont {Zhang}}, \bibinfo {author}
  {\bibfnamefont {Y.}~\bibnamefont {Chen}}, \bibinfo {author} {\bibfnamefont
  {L.}~\bibnamefont {Yang}}, \bibinfo {author} {\bibfnamefont {M.}~\bibnamefont
  {Nicklas}}, \bibinfo {author} {\bibfnamefont {F.}~\bibnamefont {Steglich}},
  and\ \bibinfo {author} {\bibfnamefont {H.~Q.}\ \bibnamefont {Yuan}},\
  }\bibfield  {title} {\bibinfo {title} {Evidence for two-gap superconductivity
  in the non-centrosymmetric compound ${{\mathrm{LaNiC}}_{2}}$},\ }\href
  {http://stacks.iop.org/1367-2630/15/i=5/a=053005} {\bibfield  {journal}
  {\bibinfo  {journal} {New J. Phys.}\ }\textbf {\bibinfo {volume} {15}},\
  \bibinfo {pages} {053005} (\bibinfo {year} {2013})}\BibitemShut {NoStop}%
\bibitem [{\citenamefont {Hillier}\ \emph {et~al.}(2012)\citenamefont
  {Hillier}, \citenamefont {Quintanilla}, \citenamefont {Mazidian},
  \citenamefont {Annett},\ and\ \citenamefont {Cywinski}}]{Hillier2012}%
  \BibitemOpen
  \bibfield  {author} {\bibinfo {author} {\bibfnamefont {A.~D.}\ \bibnamefont
  {Hillier}}, \bibinfo {author} {\bibfnamefont {J.}~\bibnamefont
  {Quintanilla}}, \bibinfo {author} {\bibfnamefont {B.}~\bibnamefont
  {Mazidian}}, \bibinfo {author} {\bibfnamefont {J.~F.}\ \bibnamefont
  {Annett}}, and\ \bibinfo {author} {\bibfnamefont {R.}~\bibnamefont
  {Cywinski}},\ }\bibfield  {title} {\bibinfo {title} {Nonunitary triplet
  pairing in the centrosymmetric superconductor {${\mathrm{LaNiGa}}_{2}$}},\
  }\href {https://doi.org/10.1103/PhysRevLett.109.097001} {\bibfield  {journal}
  {\bibinfo  {journal} {Phys. Rev. Lett.}\ }\textbf {\bibinfo {volume} {109}},\
  \bibinfo {pages} {097001} (\bibinfo {year} {2012})}\BibitemShut {NoStop}%
\bibitem [{\citenamefont {Weng}\ \emph {et~al.}(2016)\citenamefont {Weng},
  \citenamefont {Zhang}, \citenamefont {Smidman}, \citenamefont {Shang},
  \citenamefont {Quintanilla}, \citenamefont {Annett}, \citenamefont {Nicklas},
  \citenamefont {Pang}, \citenamefont {Jiao}, \citenamefont {Jiang},
  \citenamefont {Chen}, \citenamefont {Steglich},\ and\ \citenamefont
  {Yuan}}]{Weng2016}%
  \BibitemOpen
  \bibfield  {author} {\bibinfo {author} {\bibfnamefont {Z.~F.}\ \bibnamefont
  {Weng}}, \bibinfo {author} {\bibfnamefont {J.~L.}\ \bibnamefont {Zhang}},
  \bibinfo {author} {\bibfnamefont {M.}~\bibnamefont {Smidman}}, \bibinfo
  {author} {\bibfnamefont {T.}~\bibnamefont {Shang}}, \bibinfo {author}
  {\bibfnamefont {J.}~\bibnamefont {Quintanilla}}, \bibinfo {author}
  {\bibfnamefont {J.~F.}\ \bibnamefont {Annett}}, \bibinfo {author}
  {\bibfnamefont {M.}~\bibnamefont {Nicklas}}, \bibinfo {author} {\bibfnamefont
  {G.~M.}\ \bibnamefont {Pang}}, \bibinfo {author} {\bibfnamefont
  {L.}~\bibnamefont {Jiao}}, \bibinfo {author} {\bibfnamefont {W.~B.}\
  \bibnamefont {Jiang}}, \bibinfo {author} {\bibfnamefont {Y.}~\bibnamefont
  {Chen}}, \bibinfo {author} {\bibfnamefont {F.}~\bibnamefont {Steglich}}, and\
  \bibinfo {author} {\bibfnamefont {H.~Q.}\ \bibnamefont {Yuan}},\ }\bibfield
  {title} {\bibinfo {title} {Two-gap superconductivity in
  ${{\mathrm{LaNiGa}}_{2}}$ with nonunitary triplet pairing and even parity gap
  symmetry},\ }\href {https://doi.org/10.1103/PhysRevLett.117.027001}
  {\bibfield  {journal} {\bibinfo  {journal} {Phys. Rev. Lett.}\ }\textbf
  {\bibinfo {volume} {117}},\ \bibinfo {pages} {027001} (\bibinfo {year}
  {2016})}\BibitemShut {NoStop}%
\bibitem [{\citenamefont {Ghosh}\ \emph
  {et~al.}(2020{\natexlab{b}})\citenamefont {Ghosh}, \citenamefont {Csire},
  \citenamefont {Whittlesea}, \citenamefont {Annett}, \citenamefont {Gradhand},
  \citenamefont {\'Ujfalussy},\ and\ \citenamefont {Quintanilla}}]{Ghosh2019}%
  \BibitemOpen
  \bibfield  {author} {\bibinfo {author} {\bibfnamefont {S.~K.}\ \bibnamefont
  {Ghosh}}, \bibinfo {author} {\bibfnamefont {G.}~\bibnamefont {Csire}},
  \bibinfo {author} {\bibfnamefont {P.}~\bibnamefont {Whittlesea}}, \bibinfo
  {author} {\bibfnamefont {J.~F.}\ \bibnamefont {Annett}}, \bibinfo {author}
  {\bibfnamefont {M.}~\bibnamefont {Gradhand}}, \bibinfo {author}
  {\bibfnamefont {B.}~\bibnamefont {\'Ujfalussy}}, and\ \bibinfo {author}
  {\bibfnamefont {J.}~\bibnamefont {Quintanilla}},\ }\bibfield  {title}
  {\bibinfo {title} {Quantitative theory of triplet pairing in the
  unconventional superconductor {${\mathrm{LaNiGa}}_{2}$}},\ }\href
  {https://doi.org/10.1103/PhysRevB.101.100506} {\bibfield  {journal} {\bibinfo
   {journal} {Phys. Rev. B}\ }\textbf {\bibinfo {volume} {101}},\ \bibinfo
  {pages} {100506(R)} (\bibinfo {year} {2020}{\natexlab{b}})}\BibitemShut
  {NoStop}%
\bibitem [{\citenamefont {Miraglia}\ \emph {et~al.}(1987)\citenamefont
  {Miraglia}, \citenamefont {Hodeau}, \citenamefont {De~Bergevin},
  \citenamefont {Marezio},\ and\ \citenamefont {Espinosa}}]{Miraglia1987}%
  \BibitemOpen
  \bibfield  {author} {\bibinfo {author} {\bibfnamefont {S.}~\bibnamefont
  {Miraglia}}, \bibinfo {author} {\bibfnamefont {J.~L.}\ \bibnamefont
  {Hodeau}}, \bibinfo {author} {\bibfnamefont {F.}~\bibnamefont {De~Bergevin}},
  \bibinfo {author} {\bibfnamefont {M.}~\bibnamefont {Marezio}}, and\ \bibinfo
  {author} {\bibfnamefont {G.~P.}\ \bibnamefont {Espinosa}},\ }\bibfield
  {title} {\bibinfo {title} {Structural studies by electron and x-ray
  diffraction of the disordered phases
  {II':(Sn$_{1-x}$Tb$_x$)Tb$_4$Rh$_6$Sn$_{18}$ and
  (Sn$_{1-x}$Dy$_x$)Dy$_4$Os$_6$Sn$_{18}$}},\ }\href
  {https://doi.org/10.1107/S0108768187098276} {\bibfield  {journal} {\bibinfo
  {journal} {Acta Crystallogr. B}\ }\textbf {\bibinfo {volume} {43}},\ \bibinfo
  {pages} {76} (\bibinfo {year} {1987})}\BibitemShut {NoStop}%
\bibitem [{\citenamefont {Bhattacharyya}\ \emph
  {et~al.}(2015{\natexlab{a}})\citenamefont {Bhattacharyya}, \citenamefont
  {Adroja}, \citenamefont {Quintanilla}, \citenamefont {Hillier}, \citenamefont
  {Kase}, \citenamefont {Strydom},\ and\ \citenamefont
  {Akimitsu}}]{Bhattacharyya2015}%
  \BibitemOpen
  \bibfield  {author} {\bibinfo {author} {\bibfnamefont {A.}~\bibnamefont
  {Bhattacharyya}}, \bibinfo {author} {\bibfnamefont {D.~T.}\ \bibnamefont
  {Adroja}}, \bibinfo {author} {\bibfnamefont {J.}~\bibnamefont {Quintanilla}},
  \bibinfo {author} {\bibfnamefont {A.~D.}\ \bibnamefont {Hillier}}, \bibinfo
  {author} {\bibfnamefont {N.}~\bibnamefont {Kase}}, \bibinfo {author}
  {\bibfnamefont {A.~M.}\ \bibnamefont {Strydom}}, and\ \bibinfo {author}
  {\bibfnamefont {J.}~\bibnamefont {Akimitsu}},\ }\bibfield  {title} {\bibinfo
  {title} {Broken time-reversal symmetry probed by muon spin relaxation in the
  caged type superconductor
  {${\mathrm{Lu}}_{5}{\mathrm{Rh}}_{6}{\mathrm{Sn}}_{18}$}},\ }\href
  {https://doi.org/10.1103/PhysRevB.91.060503} {\bibfield  {journal} {\bibinfo
  {journal} {Phys. Rev. B}\ }\textbf {\bibinfo {volume} {91}},\ \bibinfo
  {pages} {060503(R)} (\bibinfo {year} {2015}{\natexlab{a}})}\BibitemShut
  {NoStop}%
\bibitem [{\citenamefont {Bhattacharyya}\ \emph
  {et~al.}(2015{\natexlab{b}})\citenamefont {Bhattacharyya}, \citenamefont
  {Adroja}, \citenamefont {Kase}, \citenamefont {Hillier}, \citenamefont
  {Akimitsu},\ and\ \citenamefont {Strydom}}]{Bhattacharyya2015A}%
  \BibitemOpen
  \bibfield  {author} {\bibinfo {author} {\bibfnamefont {A.}~\bibnamefont
  {Bhattacharyya}}, \bibinfo {author} {\bibfnamefont {D.~T.}\ \bibnamefont
  {Adroja}}, \bibinfo {author} {\bibfnamefont {N.}~\bibnamefont {Kase}},
  \bibinfo {author} {\bibfnamefont {A.~D.}\ \bibnamefont {Hillier}}, \bibinfo
  {author} {\bibfnamefont {J.}~\bibnamefont {Akimitsu}}, and\ \bibinfo {author}
  {\bibfnamefont {A.}~\bibnamefont {Strydom}},\ }\bibfield  {title} {\bibinfo
  {title} {Unconventional superconductivity in
  {${\mathbf{Y}}_{\mathbf{5}}{\mathbf{Rh}}_{\mathbf{6}}{\mathbf{Sn}}_{\mathbf{18}}$}
  probed by muon spin relaxation},\ }\href {https://doi.org/10.1038/srep12926}
  {\bibfield  {journal} {\bibinfo  {journal} {Scientific Reports}\ }\textbf
  {\bibinfo {volume} {5}},\ \bibinfo {pages} {12926} (\bibinfo {year}
  {2015}{\natexlab{b}})}\BibitemShut {NoStop}%
\bibitem [{\citenamefont {Bhattacharyya}\ \emph {et~al.}(2018)\citenamefont
  {Bhattacharyya}, \citenamefont {Adroja}, \citenamefont {Kase}, \citenamefont
  {Hillier}, \citenamefont {Strydom},\ and\ \citenamefont
  {Akimitsu}}]{Bhattacharyya2018}%
  \BibitemOpen
  \bibfield  {author} {\bibinfo {author} {\bibfnamefont {A.}~\bibnamefont
  {Bhattacharyya}}, \bibinfo {author} {\bibfnamefont {D.~T.}\ \bibnamefont
  {Adroja}}, \bibinfo {author} {\bibfnamefont {N.}~\bibnamefont {Kase}},
  \bibinfo {author} {\bibfnamefont {A.~D.}\ \bibnamefont {Hillier}}, \bibinfo
  {author} {\bibfnamefont {A.~M.}\ \bibnamefont {Strydom}}, and\ \bibinfo
  {author} {\bibfnamefont {J.}~\bibnamefont {Akimitsu}},\ }\bibfield  {title}
  {\bibinfo {title} {Unconventional superconductivity in the cage-type compound
  {${\mathbf{Sc}}_{\mathbf{5}}{\mathbf{Rh}}_{\mathbf{6}}{\mathbf{Sn}}_{\mathbf{18}}$}},\
  }\href {https://doi.org/10.1103/PhysRevB.98.024511} {\bibfield  {journal}
  {\bibinfo  {journal} {Phys. Rev. B}\ }\textbf {\bibinfo {volume} {98}},\
  \bibinfo {pages} {024511} (\bibinfo {year} {2018})}\BibitemShut {NoStop}%
\bibitem [{\citenamefont {Feig}\ \emph {et~al.}(2020)\citenamefont {Feig},
  \citenamefont {Schnelle}, \citenamefont {Maisuradze}, \citenamefont {Amon},
  \citenamefont {Baines}, \citenamefont {Nicklas}, \citenamefont {Seiro},
  \citenamefont {Howald}, \citenamefont {Khasanov}, \citenamefont
  {Leithe-Jasper},\ and\ \citenamefont {Gumeniuk}}]{Feig2020}%
  \BibitemOpen
  \bibfield  {author} {\bibinfo {author} {\bibfnamefont {M.}~\bibnamefont
  {Feig}}, \bibinfo {author} {\bibfnamefont {W.}~\bibnamefont {Schnelle}},
  \bibinfo {author} {\bibfnamefont {A.}~\bibnamefont {Maisuradze}}, \bibinfo
  {author} {\bibfnamefont {A.}~\bibnamefont {Amon}}, \bibinfo {author}
  {\bibfnamefont {C.}~\bibnamefont {Baines}}, \bibinfo {author} {\bibfnamefont
  {M.}~\bibnamefont {Nicklas}}, \bibinfo {author} {\bibfnamefont
  {S.}~\bibnamefont {Seiro}}, \bibinfo {author} {\bibfnamefont
  {L.}~\bibnamefont {Howald}}, \bibinfo {author} {\bibfnamefont
  {R.}~\bibnamefont {Khasanov}}, \bibinfo {author} {\bibfnamefont
  {A.}~\bibnamefont {Leithe-Jasper}}, and\ \bibinfo {author} {\bibfnamefont
  {R.}~\bibnamefont {Gumeniuk}},\ }\bibfield  {title} {\bibinfo {title}
  {Conventional isotropic $s$-wave superconductivity with strong
  electron-phonon coupling in
  {${\mathrm{Sc}}_{5}{\mathrm{Rh}}_{6}{\mathrm{Sn}}_{18}$}},\ }\href
  {https://doi.org/10.1103/PhysRevB.102.024508} {\bibfield  {journal} {\bibinfo
   {journal} {Phys. Rev. B}\ }\textbf {\bibinfo {volume} {102}},\ \bibinfo
  {pages} {024508} (\bibinfo {year} {2020})}\BibitemShut {NoStop}%
\bibitem [{\citenamefont {Kase}\ \emph {et~al.}(2011)\citenamefont {Kase},
  \citenamefont {Inoue}, \citenamefont {Hayamizu},\ and\ \citenamefont
  {Akimitsu}}]{Kase2011}%
  \BibitemOpen
  \bibfield  {author} {\bibinfo {author} {\bibfnamefont {N.}~\bibnamefont
  {Kase}}, \bibinfo {author} {\bibfnamefont {K.}~\bibnamefont {Inoue}},
  \bibinfo {author} {\bibfnamefont {H.}~\bibnamefont {Hayamizu}}, and\ \bibinfo
  {author} {\bibfnamefont {J.}~\bibnamefont {Akimitsu}},\ }\bibfield  {title}
  {\bibinfo {title} {Highly anisotropic gap function in a nonmagnetic
  superconductor {Y$_5$Rh$_6$Sn$_{18}$}},\ }\href
  {https://doi.org/10.1143/JPSJS.80SA.SA112} {\bibfield  {journal} {\bibinfo
  {journal} {J. Phys. Soc. Jpn.}\ }\textbf {\bibinfo {volume} {80}},\ \bibinfo
  {pages} {SA112} (\bibinfo {year} {2011})}\BibitemShut {NoStop}%
\bibitem [{\citenamefont {Kase}\ \emph {et~al.}(2012)\citenamefont {Kase},
  \citenamefont {Kittaka}, \citenamefont {Sakakibara},\ and\ \citenamefont
  {Akimitsu}}]{Kase2011b}%
  \BibitemOpen
  \bibfield  {author} {\bibinfo {author} {\bibfnamefont {N.}~\bibnamefont
  {Kase}}, \bibinfo {author} {\bibfnamefont {S.}~\bibnamefont {Kittaka}},
  \bibinfo {author} {\bibfnamefont {T.}~\bibnamefont {Sakakibara}}, and\
  \bibinfo {author} {\bibfnamefont {J.}~\bibnamefont {Akimitsu}},\ }\bibfield
  {title} {\bibinfo {title} {Superconducting gap structure of the cage compound
  {Sc$_5$Rh$_6$Sn$_{18}$}},\ }\href {https://doi.org/10.1143/JPSJS.81SB.SB016}
  {\bibfield  {journal} {\bibinfo  {journal} {J. Phys. Soc. Jpn.}\ }\textbf
  {\bibinfo {volume} {81}},\ \bibinfo {pages} {SB016} (\bibinfo {year}
  {2012})}\BibitemShut {NoStop}%
\bibitem [{\citenamefont {Zhang}\ \emph {et~al.}(2015)\citenamefont {Zhang},
  \citenamefont {Xu}, \citenamefont {Kuo}, \citenamefont {Hong}, \citenamefont
  {Wang}, \citenamefont {Cai}, \citenamefont {Dong}, \citenamefont {Lue},\ and\
  \citenamefont {Li}}]{Zhang2015a}%
  \BibitemOpen
  \bibfield  {author} {\bibinfo {author} {\bibfnamefont {Z.}~\bibnamefont
  {Zhang}}, \bibinfo {author} {\bibfnamefont {Y.}~\bibnamefont {Xu}}, \bibinfo
  {author} {\bibfnamefont {C.~N.}\ \bibnamefont {Kuo}}, \bibinfo {author}
  {\bibfnamefont {X.~C.}\ \bibnamefont {Hong}}, \bibinfo {author}
  {\bibfnamefont {M.~X.}\ \bibnamefont {Wang}}, \bibinfo {author}
  {\bibfnamefont {P.~L.}\ \bibnamefont {Cai}}, \bibinfo {author} {\bibfnamefont
  {J.~K.}\ \bibnamefont {Dong}}, \bibinfo {author} {\bibfnamefont {C.~S.}\
  \bibnamefont {Lue}}, and\ \bibinfo {author} {\bibfnamefont {S.~Y.}\
  \bibnamefont {Li}},\ }\bibfield  {title} {\bibinfo {title} {Nodeless
  superconducting gap in the caged-type superconductors {Y$_5$Rh$_6$Sn$_{18}$
  and Lu$_5$Rh$_6$Sn$_{18}$}},\ }\href
  {https://doi.org/10.1088/0953-2048/28/10/105008} {\bibfield  {journal}
  {\bibinfo  {journal} {Supercond. Sci. Technol.}\ }\textbf {\bibinfo {volume}
  {28}},\ \bibinfo {pages} {105008} (\bibinfo {year} {2015})}\BibitemShut
  {NoStop}%
\bibitem [{\citenamefont {Remeika}\ \emph {et~al.}(1980)\citenamefont
  {Remeika}, \citenamefont {Espinosa}, \citenamefont {Cooper}, \citenamefont
  {Barz}, \citenamefont {Rowell}, \citenamefont {McWhan}, \citenamefont
  {Vandenberg}, \citenamefont {Moncton}, \citenamefont {Fisk}, \citenamefont
  {Woolf}, \citenamefont {Hamaker}, \citenamefont {Maple}, \citenamefont
  {Shirane},\ and\ \citenamefont {Thomlinson}}]{Remeika1980}%
  \BibitemOpen
  \bibfield  {author} {\bibinfo {author} {\bibfnamefont {J.}~\bibnamefont
  {Remeika}}, \bibinfo {author} {\bibfnamefont {G.}~\bibnamefont {Espinosa}},
  \bibinfo {author} {\bibfnamefont {A.}~\bibnamefont {Cooper}}, \bibinfo
  {author} {\bibfnamefont {H.}~\bibnamefont {Barz}}, \bibinfo {author}
  {\bibfnamefont {J.}~\bibnamefont {Rowell}}, \bibinfo {author} {\bibfnamefont
  {D.}~\bibnamefont {McWhan}}, \bibinfo {author} {\bibfnamefont
  {J.}~\bibnamefont {Vandenberg}}, \bibinfo {author} {\bibfnamefont
  {D.}~\bibnamefont {Moncton}}, \bibinfo {author} {\bibfnamefont
  {Z.}~\bibnamefont {Fisk}}, \bibinfo {author} {\bibfnamefont {L.}~\bibnamefont
  {Woolf}}, \bibinfo {author} {\bibfnamefont {H.}~\bibnamefont {Hamaker}},
  \bibinfo {author} {\bibfnamefont {M.}~\bibnamefont {Maple}}, \bibinfo
  {author} {\bibfnamefont {G.}~\bibnamefont {Shirane}}, and\ \bibinfo {author}
  {\bibfnamefont {W.}~\bibnamefont {Thomlinson}},\ }\bibfield  {title}
  {\bibinfo {title} {A new family of ternary intermetallic
  superconducting/magnetic stannides},\ }\href
  {https://doi.org/https://doi.org/10.1016/0038-1098(80)91099-6} {\bibfield
  {journal} {\bibinfo  {journal} {Solid State Commun.}\ }\textbf {\bibinfo
  {volume} {34}},\ \bibinfo {pages} {923 } (\bibinfo {year}
  {1980})}\BibitemShut {NoStop}%
  \bibitem [{\citenamefont {Keen}\ \emph {et~al.}(2006)\citenamefont {Keen},
  \citenamefont {Gutmann},\ and\ \citenamefont {Wilson}}]{Keen2006}%
  \BibitemOpen
  \bibfield  {author} {\bibinfo {author} {\bibfnamefont {D.~A.}\ \bibnamefont
  {Keen}}, \bibinfo {author} {\bibfnamefont {M.~J.}\ \bibnamefont {Gutmann}},
  and\ \bibinfo {author} {\bibfnamefont {C.~C.}\ \bibnamefont {Wilson}},\
  }\bibfield  {title} {\bibinfo {title} {{{SXD} {--} the single-crystal
  diffractometer at the {ISIS} spallation neutron source}},\ }\href
  {https://doi.org/10.1107/S0021889806025921} {\bibfield  {journal} {\bibinfo
  {journal} {J. Appl. Crystallogr.}\ }\textbf {\bibinfo {volume} {39}},\
  \bibinfo {pages} {714} (\bibinfo {year} {2006})}\BibitemShut {NoStop}%
\bibitem [{\citenamefont {Prozorov}\ \emph {et~al.}(2000)\citenamefont
  {Prozorov}, \citenamefont {Giannetta}, \citenamefont {Carrington},\ and\
  \citenamefont {Araujo-Moreira}}]{Gfactor}%
  \BibitemOpen
  \bibfield  {author} {\bibinfo {author} {\bibfnamefont {R.}~\bibnamefont
  {Prozorov}}, \bibinfo {author} {\bibfnamefont {R.~W.}\ \bibnamefont
  {Giannetta}}, \bibinfo {author} {\bibfnamefont {A.}~\bibnamefont
  {Carrington}}, and\ \bibinfo {author} {\bibfnamefont {F.~M.}\ \bibnamefont
  {Araujo-Moreira}},\ }\bibfield  {title} {\bibinfo {title} {Meissner-london
  state in superconductors of rectangular cross section in a perpendicular
  magnetic field},\ }\href {https://doi.org/10.1103/PhysRevB.62.115} {\bibfield
   {journal} {\bibinfo  {journal} {Phys. Rev. B}\ }\textbf {\bibinfo {volume}
  {62}},\ \bibinfo {pages} {115} (\bibinfo {year} {2000})}\BibitemShut
  {NoStop}%
\bibitem [{\citenamefont {Kresse}\ and\ \citenamefont {Hafner}(1993)}]{VASP01}%
  \BibitemOpen
  \bibfield  {author} {\bibinfo {author} {\bibfnamefont {G.}~\bibnamefont
  {Kresse}} and\ \bibinfo {author} {\bibfnamefont {J.}~\bibnamefont {Hafner}},\
  }\bibfield  {title} {\bibinfo {title} {Ab initio molecular dynamics for
  liquid metals},\ }\href {https://doi.org/10.1103/PhysRevB.47.558} {\bibfield
  {journal} {\bibinfo  {journal} {Phys. Rev. B}\ }\textbf {\bibinfo {volume}
  {47}},\ \bibinfo {pages} {558} (\bibinfo {year} {1993})}\BibitemShut
  {NoStop}%
\bibitem [{\citenamefont {Kresse}\ and\ \citenamefont
  {Joubert}(1999)}]{VASP02}%
  \BibitemOpen
  \bibfield  {author} {\bibinfo {author} {\bibfnamefont {G.}~\bibnamefont
  {Kresse}} and\ \bibinfo {author} {\bibfnamefont {D.}~\bibnamefont {Joubert}},\
  }\bibfield  {title} {\bibinfo {title} {From ultrasoft pseudopotentials to the
  projector augmented-wave method},\ }\href
  {https://doi.org/10.1103/PhysRevB.59.1758} {\bibfield  {journal} {\bibinfo
  {journal} {Phys. Rev. B}\ }\textbf {\bibinfo {volume} {59}},\ \bibinfo
  {pages} {1758} (\bibinfo {year} {1999})}\BibitemShut {NoStop}%
\bibitem [{\citenamefont {Perdew}\ \emph {et~al.}(1996)\citenamefont {Perdew},
  \citenamefont {Burke},\ and\ \citenamefont {Ernzerhof}}]{GGA}%
  \BibitemOpen
  \bibfield  {author} {\bibinfo {author} {\bibfnamefont {J.~P.}\ \bibnamefont
  {Perdew}}, \bibinfo {author} {\bibfnamefont {K.}~\bibnamefont {Burke}}, and\
  \bibinfo {author} {\bibfnamefont {M.}~\bibnamefont {Ernzerhof}},\ }\bibfield
  {title} {\bibinfo {title} {Generalized gradient approximation made simple},\
  }\href {https://doi.org/10.1103/PhysRevLett.77.3865} {\bibfield  {journal}
  {\bibinfo  {journal} {Phys. Rev. Lett.}\ }\textbf {\bibinfo {volume} {77}},\
  \bibinfo {pages} {3865} (\bibinfo {year} {1996})}\BibitemShut {NoStop}%
\bibitem [{\citenamefont {Orlando}\ \emph {et~al.}(1979)\citenamefont
  {Orlando}, \citenamefont {McNiff}, \citenamefont {Foner},\ and\ \citenamefont
  {Beasley}}]{Orlando1979}%
  \BibitemOpen
  \bibfield  {author} {\bibinfo {author} {\bibfnamefont {T.~P.}\ \bibnamefont
  {Orlando}}, \bibinfo {author} {\bibfnamefont {E.~J.}\ \bibnamefont {McNiff}},
  \bibinfo {author} {\bibfnamefont {S.}~\bibnamefont {Foner}}, and\ \bibinfo
  {author} {\bibfnamefont {M.~R.}\ \bibnamefont {Beasley}},\ }\bibfield
  {title} {\bibinfo {title} {Critical fields, {P}auli paramagnetic limiting,
  and material parameters of {${\mathrm{Nb}}_{3}$Sn and
  ${\mathrm{V}}_{3}$Si}},\ }\href {https://doi.org/10.1103/PhysRevB.19.4545}
  {\bibfield  {journal} {\bibinfo  {journal} {Phys. Rev. B}\ }\textbf {\bibinfo
  {volume} {19}},\ \bibinfo {pages} {4545} (\bibinfo {year}
  {1979})}\BibitemShut {NoStop}%
\bibitem [{\citenamefont {Prozorov}\ and\ \citenamefont
  {Giannetta}(2006)}]{Prozorov2006}%
  \BibitemOpen
  \bibfield  {author} {\bibinfo {author} {\bibfnamefont {R.}~\bibnamefont
  {Prozorov}} and\ \bibinfo {author} {\bibfnamefont {R.~W.}\ \bibnamefont
  {Giannetta}},\ }\bibfield  {title} {\bibinfo {title} {Magnetic penetration
  depth in unconventional superconductors},\ }\href
  {http://stacks.iop.org/0953-2048/19/i=8/a=R01} {\bibfield  {journal}
  {\bibinfo  {journal} {Superconductor Science and Technology}\ }\textbf
  {\bibinfo {volume} {19}},\ \bibinfo {pages} {R41} (\bibinfo {year}
  {2006})}\BibitemShut {NoStop}%
\bibitem [{\citenamefont {Maisuradze}\ \emph {et~al.}(2009)\citenamefont
  {Maisuradze}, \citenamefont {Nicklas}, \citenamefont {Gumeniuk},
  \citenamefont {Baines}, \citenamefont {Schnelle}, \citenamefont {Rosner},
  \citenamefont {Leithe-Jasper}, \citenamefont {Grin},\ and\ \citenamefont
  {Khasanov}}]{Maisuradze2009}%
  \BibitemOpen
  \bibfield  {author} {\bibinfo {author} {\bibfnamefont {A.}~\bibnamefont
  {Maisuradze}}, \bibinfo {author} {\bibfnamefont {M.}~\bibnamefont {Nicklas}},
  \bibinfo {author} {\bibfnamefont {R.}~\bibnamefont {Gumeniuk}}, \bibinfo
  {author} {\bibfnamefont {C.}~\bibnamefont {Baines}}, \bibinfo {author}
  {\bibfnamefont {W.}~\bibnamefont {Schnelle}}, \bibinfo {author}
  {\bibfnamefont {H.}~\bibnamefont {Rosner}}, \bibinfo {author} {\bibfnamefont
  {A.}~\bibnamefont {Leithe-Jasper}}, \bibinfo {author} {\bibfnamefont
  {Y.}~\bibnamefont {Grin}}, and\ \bibinfo {author} {\bibfnamefont
  {R.}~\bibnamefont {Khasanov}},\ }\bibfield  {title} {\bibinfo {title}
  {Superfluid density and energy gap function of superconducting
  {${\mathrm{PrPt}}_{4}{\mathrm{Ge}}_{12}$}},\ }\href
  {https://doi.org/10.1103/PhysRevLett.103.147002} {\bibfield  {journal}
  {\bibinfo  {journal} {Phys. Rev. Lett.}\ }\textbf {\bibinfo {volume} {103}},\
  \bibinfo {pages} {147002} (\bibinfo {year} {2009})}\BibitemShut {NoStop}%
\bibitem [{\citenamefont {Tinkham}(2004)}]{Tinkham2004}%
  \BibitemOpen
  \bibfield  {author} {\bibinfo {author} {\bibfnamefont {M.}~\bibnamefont
  {Tinkham}},\ }\href@noop {} {\emph {\bibinfo {title} {Introduction to
  superconductivity}}}\ (\bibinfo  {publisher} {Courier Corporation},\ \bibinfo
  {year} {2004})\BibitemShut {NoStop}%
\bibitem [{\citenamefont {Xia}\ \emph {et~al.}(2006)\citenamefont {Xia},
  \citenamefont {Maeno}, \citenamefont {Beyersdorf}, \citenamefont {Fejer},\
  and\ \citenamefont {Kapitulnik}}]{Xia2006}%
  \BibitemOpen
  \bibfield  {author} {\bibinfo {author} {\bibfnamefont {J.}~\bibnamefont
  {Xia}}, \bibinfo {author} {\bibfnamefont {Y.}~\bibnamefont {Maeno}}, \bibinfo
  {author} {\bibfnamefont {P.~T.}\ \bibnamefont {Beyersdorf}}, \bibinfo
  {author} {\bibfnamefont {M.~M.}\ \bibnamefont {Fejer}}, and\ \bibinfo
  {author} {\bibfnamefont {A.}~\bibnamefont {Kapitulnik}},\ }\bibfield  {title}
  {\bibinfo {title} {High resolution polar {K}err effect measurements of
  {${\mathrm{Sr}}_{2}{\mathrm{RuO}}_{4}$}: {E}vidence for broken time-reversal
  symmetry in the superconducting state},\ }\href
  {https://doi.org/10.1103/PhysRevLett.97.167002} {\bibfield  {journal}
  {\bibinfo  {journal} {Phys. Rev. Lett.}\ }\textbf {\bibinfo {volume} {97}},\
  \bibinfo {pages} {167002} (\bibinfo {year} {2006})}\BibitemShut {NoStop}%
\bibitem [{\citenamefont {{Ghosh}}\ \emph {et~al.}(2018)\citenamefont
  {{Ghosh}}, \citenamefont {{Annett}},\ and\ \citenamefont
  {{Quintanilla}}}]{Ghosh2018}%
  \BibitemOpen
  \bibfield  {author} {\bibinfo {author} {\bibfnamefont {S.~K.}\ \bibnamefont
  {{Ghosh}}}, \bibinfo {author} {\bibfnamefont {J.~F.}\ \bibnamefont
  {{Annett}}}, and\ \bibinfo {author} {\bibfnamefont {J.}~\bibnamefont
  {{Quintanilla}}},\ }\bibfield  {title} {\bibinfo {title} {{Time-reversal
  symmetry breaking in superconductors through loop super-current order}},\
  }\href@noop {} {\bibfield  {journal} {\bibinfo  {journal} {arXiv e-prints}\
  ,\ \bibinfo {eid} {arXiv:1803.02618}} (\bibinfo {year} {2018})},\ \Eprint
  {https://arxiv.org/abs/1803.02618} {arXiv:1803.02618 [cond-mat.supr-con]}
  \BibitemShut {NoStop}%
\end{thebibliography}
\end{document}